\documentclass[aps,prx,twocolumn,amsmath,amssymb,nofootinbib,10pt]{revtex4}
\usepackage{graphicx}
\usepackage{amsmath,amssymb,bm} 
\usepackage{ulem}
\usepackage{soul}
\usepackage{upgreek}
\usepackage[colorlinks, linkcolor=blue]{hyperref}
\usepackage{color} 
\usepackage[papersize={8.5in,11in}]{geometry}
\usepackage{verbatim}
\usepackage{esint}
\hypersetup{
    unicode=false,          
    pdftoolbar=true,        
    pdfmenubar=true,        
    pdffitwindow=false,     
    pdfstartview={FitH},    
    pdftitle={Quantum butterfly effect in weakly interacting diffusive metals},    
    pdfauthor={Aavishkar A. Patel},     
    pdfsubject={},   
    pdfcreator={},   
    pdfproducer={}, 
    pdfkeywords={} {} {}, 
    pdfnewwindow=true,      
    colorlinks=true,       
    linkcolor=magenta, 
    citecolor=blue,        
    filecolor=magenta,      
    urlcolor=blue           
} 

\renewcommand{\vec}[1]{\boldsymbol{#1}}

\def \p {{\vec p}}
\def \a {{\vec a}}

\def \B {{\vec B}}

\def \A{{\vec A}}

\def \be {\begin{eqnarray}}
\def \ee {\end{eqnarray}}
\def \tn {\textnormal}

\begin{document}
\title{Quantum oscillations in insulators with neutral Fermi surfaces}

\author{Inti Sodemann}
\affiliation{Department of Physics, Massachusetts Institute of Technology, Cambridge, Massachusetts 02139, USA}
\author{Debanjan Chowdhury}
\affiliation{Department of Physics, Massachusetts Institute of Technology, Cambridge, Massachusetts 02139, USA}
\author{T. Senthil}
\affiliation{Department of Physics, Massachusetts Institute of Technology, Cambridge, Massachusetts 02139, USA}

\date{\today}

\begin{abstract}
We develop a theory of quantum oscillations in insulators with an emergent fermi sea of neutral fermions minimally coupled to an emergent $U(1)$ gauge field. As pointed out by Motrunich~\cite{Motrunich}, in the presence of a physical magnetic field the emergent magnetic field develops a non-zero value leading to Landau quantization for the neutral fermions. We focus on the magnetic field and temperature dependence of the analogue of the de Haas-van Alphen effect in two- and three-dimensions. At temperatures above the effective cyclotron energy, the magnetization oscillations behave similarly to those of an ordinary metal, albeit in a field of a strength that differs from the physical magnetic field. At low temperatures the oscillations evolve into a series of phase transitions. We provide analytical expressions for the amplitude and period of the oscillations in both of these regimes and simple extrapolations that capture well their crossover. We also describe oscillations in the electrical resistivity of these systems that are expected to be superimposed with the activated temperature behavior characteristic of their insulating nature and discuss suitable experimental conditions for the observation of these effects in mixed-valence insulators and triangular lattice organic materials.

\end{abstract}

\maketitle

\section{Introduction}

In the past few years a number of materials that are in close proximity to the metal to Mott insulator transition have come to the forefront as prime candidates to harbor spin liquid phases, notably the triangular lattice organic materials $\kappa-$(BEDT-TTF)$_2$Cu$_2$(CN)$_3$ and EtMe$_3$Sb[Pd(dmit)$_2$]$_2$ \cite{Shimizu03,Yamashita08,Yamashita10,Yamashita11}. These materials are charge insulators that lack spin order down to the lowest temperatures. Theoretically, such phases of matter can be understood to arise when the electron is splintered apart into fractionalized excitations that carry its charge and spin separately \cite{PAL05,OM05}. The precise nature of the spin liquid realized in these materials is still contended, but remarkably, the dmit compound remains a thermal conductor with a finite intercept of the ratio of heat conductivity to temperature down to the lowest measurable temperatures~\cite{Yamashita10}, in spite of displaying clear charge insulating behavior. It is believed that a good starting point to describe the phenomenology of these organic spin liquids is a state with a Fermi surface of emergent neutral spin-$1/2$ fermions (dubbed spinons)~\cite{Lee&Lee,MotrunichVariational}.

 In other fascinating recent developments, mixed valence insulators such as samarium hexaboride (SmB$_6$) have been seen to display de Haas-van Alphen (dHvA) oscillations in their magnetization in an applied external magnetic field.  
Initially these oscillations were attributed to the metallic surface state that SmB$_6$ is known to possess~\cite{Li14,Li16}. However subsequent experiments have raised the dramatic possibility that these quantum oscillations are a bulk effect in this electrical insulator~\cite{SS15,SSnew}, possibly related to other mysterious low-temperature anomalies in the thermodynamic and optical properties~\cite{Li14,SS15,Li16,flachbart,thompson,Armitage16}. Inspired by this situation, we recently described a new phase of matter - dubbed {\it composite exciton Fermi liquid} - in a mixed valence insulator  with neutral fermionic quasiparticles (coupled to a dynamical $U(1)$ gauge field) that form a Fermi surface~\cite{CSS}.  The composite exciton Fermi liquid is sharply distinct from other proposals which either posited a Fermi surface of Majorana fermions~\cite{Baskaran,ColemanSC}, nearly gapless bosonic excitons~\cite{CooperExc}, or magnetic breakdown mechanisms in inverted band insulators~\cite{CooperMB,FWMB,Kumar2017} as descriptions of the phenomena in SmB$_6$.

In conventional phases of matter that have a conserved charge, charge neutral quasiparticles are necessarily bosons. However, in the presence of fractionalization, neutral fermions can emerge and they can in turn form a Fermi surface under suitable conditions. The spinon Fermi-surface state is one such example.  Another classic example is the half-filled Landau level where a charge neutral fermion (i.e. the composite fermion\footnote{For arguments on neutrality of composite fermions see~\cite{NR94,NR98}.}) emerges, albeit in a metallic phase \cite{HLR}. The fermionic composite-exciton, recently proposed by us  for correlated mixed-valence insulators~\cite{CSS} is yet another example of this phenomenon.

Can electronic solids with a neutral Fermi surface in an external magnetic field show a de Haas-van Alphen effect? Can they display quantum oscillations in other properties (such as the resistivity at a non-zero temperature)?   
 This would be remarkable because conventional insulating phases respond in a rather innocuous way to applied magnetic fields: a magnetization that is linear in the external field usually develops typically with an opposite direction to try to screen it (diamagnetism)\footnote{We are imagining a non-magnetic insulating state.}, and the resistivity displays a smooth temperature activated behavior. Such behavior is in stark contrast to the situation in ordinary metals which at low temperatures display oscillations of magnetization and resistivity as a function of the magnetic fields with a frequency that diverges as $1/B$ at low fields, rendering the response a non-analytic function of $B$. Such behavior is a fingerprint of the non-perturbative modification in the low energy spectrum of the metal associated with Landau quantization. It would therefore be striking to realize insulating phases of matter displaying such quantum oscillations at weak magnetic fields, which are often thought to be fingerprints of metallic behavior.

The possibility of observing quantum oscillations in a Mott-insulator with a spinon Fermi-surface was first studied in a pioneering work by Motrunich~\cite{Motrunich}, in the context of the organic material $\kappa$-(ET)$_2$Cu$_2$(CN)$_3$. Motrunich emphasized that generically in the presence of an external magnetic field, an internal magnetic field of the emergent gauge field will develop leading to Landau quantization of the spinons. Within this model, Motrunich found, quite remarkably, that the strength of such an effective magnetic field experienced by the spinons could even be larger than the one experienced by bare electrons\footnote{The strength can be defined operationally in a gauge invariant fashion by considering the amount of Aharonov-Bohm phase that the spinon acquires in a loop enclosing some given small area as compared to that acquired by the electron.}. Another important property emphasized in Motrunich's work was the softening of the stiffness of the emergent magnetic field as one goes deeper into the insulating phase. As we will see, this effect makes the magnetization response of fractionalized neutral fermi seas differ qualitatively from those of metals once the temperature is lower than the effective cyclotron energy of the neutral fermions. In particular, this allows for the average value of the emergent magnetic field to self-consistently adjust itself to lower the energy. One of the concerns of Motrunich's work was that such reduced stiffness would lead to an enhanced tendency to form non-uniform states analogous to Condon domains observed in metals~\cite{abrikosov,DS,Domains99,Domains99b,gordon2003magnetic,Domains05,egorov2005condon}. As we will argue, we believe that this will not preempt the observation of quantum oscillations in neutral fermi seas in the semiclassical regime, to the same extent that it does not preempt the observation of quantum oscillations in metals. However, in practice it will change the precise shape of the oscillatory component of the magnetization as a function of the external magnetic field. We will also show that quantum oscillations will occur in the finite temperature resistivity but may be hard to observe except under certain suitable conditions that we will describe.

Our study is constructed around a minimal low energy effective field theory for the neutral fermi sea, and we have deliberately attempted to keep our results as universally applicable as possible. One of our focus, which complements that of Motrunich, is that we have developed a detailed quantitative theory of the temperature dependence of the quantum oscillations. In particular, we study how the high temperature regime, that closely resembles that of a metal, evolves into the low temperature regime previously identified by Motrunich. We will show that at low temperatures the quantum oscillations can be viewed as an infinite sequence of phase transitions between states in which the emergent magnetic field takes different values, and we will see how this tendency evolves into the more conventional form of quantum oscillations as the temperature is raised. Our results are widely applicable to fractionalized phases with fermi surfaces of neutral fermions with a charge-gap, i.e. for insulating states\footnote{They are not applicable to the half-filled Landau level which is metallic.}. These include the conventional $U(1)$ spin liquids with spinon fermi surface and the composite exciton Fermi liquid in mixed valence insulators~\cite{CSS}.

Our paper is organized as follows: In Section~\ref{setup} we discuss the general setup and principles needed to compute the magnetization of fractionalized neutral fermi seas. In Section~\ref{2D} we study quantum oscillations in the magnetization of two-dimensional fractionalized neutral fermi seas, and will show the interesting property that their period at low temperatures will in general be different from their period at higher temperatures.  In Section~\ref{3DfiniteT} we will develop the theory of quantum oscillations in three dimensional fractionalized fermi seas, such as that for the composite exciton Fermi liquid proposed to arise in mixed valence insulators~\cite{CSS}. In contrast to the two-dimensional case, the period of the oscillations in three dimensions does not change with temperature. In Section~\ref{resistivity} we will show that fractionalized fermi seas can display also a form of Shubnikov-de Haas oscillations in the resistivity at finite temperature superimposed with the activated behavior characteristic of charge insulators. We close in Section~\ref{sum} with a summary of our results and a discussion on their implications on current and future experiments both in organic spin liquids and in mixed valence insulators. We summarize the main formulas for the amplitude and period of the oscillations in Table~\ref{sumtab}.

\section{General Setup}\label{setup}
\subsection{Low energy field theory}

We are considering phases of matter which can emerge out of a Hilbert space where the local degrees of freedom are electrons that couple minimally to the static external gauge-field $A=(A_0,\A)$. We assume that the total electron number, $Q\in \mathbb{Z}$, is a good quantum number and we will refer to this quantum number as the {\it charge}. We are interested in phases with an emergent fermion with no physical charge, coupled to an emergent $U(1)$ gauge field, which we denote $a=(a_0,\a)$, and a gapped boson that carries the physical charge $Q=1$\footnote{This is one of the simplest patterns of fractionalization that allows for the emergence of a neutral fermion in a Hilbert space of microscopic electrons.}. Let us denote the neutral fermionic creation and annihilation operators by $\psi^\dagger,~\psi$ (where they satisfy the usual anticommutation algebra) and the bosonic operators by $\varphi^\dagger,~\varphi$. Suppose now that the physical electron is represented in terms of these emergent fractionalized quasiparticles as: $c^\dagger=\psi^\dagger \varphi^\dagger$.  These fractionalized particles must necessarily carry {\it gauge charge} under $a$, which we denote as $q$, because they are {\it non-local}~\footnote{Namely $\psi^\dagger$ and $b^\dagger$ cannot be written in terms of of a finite number of electron creation/annihilation operators acting over a finite region of space.}. Moreover, these excitations  carry opposite gauge charge under $a$, as demanded by locality of the physical electron. Without loss of generality, we take them to be $q_\psi=1$ and $q_\varphi=-1$. 

In the above representation, we recover the ordinary metallic phase when the boson is {\it condensed}, $\langle \varphi \rangle\neq 0$, while an insulating state with a Fermi-surface of the neutral fermion can emerge when the boson is {\it gapped}, $\langle \varphi \rangle = 0$. For concreteness, let us write down a Lagrangian describing the low energy physics as follows:

\be
\mathcal{L}&=&\psi^\dagger \left(i \partial_t-a_{0}-\frac{(\p-\a)^2}{2m_\psi}\right) \psi  \nonumber\\
&+&|(i\partial_\mu+a_\mu-A_\mu)\varphi|^2-g|\varphi|^2-\frac{u}{2}|\varphi|^4+\cdots
\ee
\noindent where $u,g,m_\psi$ are effective parameters, and any other terms that are invariant under gauge transformations of $a$ are also allowed\footnote{Strictly speaking this  Lagrangian is appropriate near the critical point associated with boson condensation at fixed boson number. Otherwise a linear in time derivative term for the boson would dominate over the relativistic quadratic term at low energies. However these details will not affect the mean field treatment that we employ here as the only requirement is that the bosons are gapped.}. Here the fermions have a finite density, and our interest is to describe the response of the system in the fractionalized phase to the presence of an external magnetic field $\B=\nabla\times \A$. All of the discussion that follows in this paper can be viewed as a mean field treatment of the physics contained within this Lagrangian.

\subsection{Thermodynamics of magnetic systems} 

Let us briefly review the thermodynamics of magnetic fields in matter for the sake of completeness. Thermodynamic quantities that are well behaved and that do not undergo spontaneous symmetry-breaking typically fall into two categories: conserved quantities and parameters of the Hamiltonian that can be macroscopically controlled. Faraday's law implies that one can consider the average physical net magnetic flux in any direction as a conserved quantity: 

\be
B_i=\frac{1}{V}\int d^d {\bf x} \ B_i({\bf x}).
\ee 

Thermodynamics of any system in the presence of magnetic fields can be described in a microcanonical magnetic ensemble, by specifying $B_i$, or in a magnetic canonical ensemble, by specifying an associated conjugate variable to $B_i$, which is $H_i$. In this paper, we will perform our calculations in the microcanical ensemble and specify $B_i$.  The conceptual advantage of viewing $B_i$ as a conserved quantity is that one can consider systems placed on strictly periodic geometries with a non-zero average $B_i$ and describe their thermodynamics without boundaries, entirely bypassing questions related to the external sources of magnetic fields and boundary currents. If $u$ is the energy of the system per unit volume (including the ``vaccum" magnetic field energy), $s$ its entropy per unit volume, then $f$, its Helmholtz free energy per unit volume, is: 

\be
f(T,\B,\{\lambda\})=u-Ts,
\ee

\noindent where $\{\lambda\}$ denotes the set of other non-magnetic thermodynamic variables. The stability of the thermodynamic state requires the following to be a positive definite matrix:

\be
\frac{\partial^2 f}{\partial B_i \partial B_j}\bigg|_{T, \{\lambda\}} .
\ee

\noindent If we use units in which the vacuum magnetic field energy density is $\B^2/2$ (i.e. the vacuum permeability is set to $1$), the physical magnetization of a system is given by:

\be
4\pi {\bf M}=\B-\frac{\partial f}{\partial \B}.
\ee

\noindent Therefore, the task of describing the static magnetic response of a system reduces to finding the form of the free energy as a function of the average magnetic field. The conjugate field $H_i$ can be obtained as $H_i=\partial f/\partial B_i$~\footnote{In experiments one can only control the external magnetic field $\mathbf{B}_0$ in which the sample is placed and this generally differs from the field $\mathbf{B}$ inside the sample. $\mathbf{B}$ will ultimately be determined by details of the sample geometry. In the case of an ellipsoidal sample one controls the following linear combination $\mathbf{B}_0=(1-n)\mathbf{H}+n \mathbf{B}$, where $n\in[0,1]$ is the demagnetization factor. Thus $\mathbf{B}_0$ interpolates between $\bf{H}$, at $n=0$, in the limit in which the ellipsoid becomes a cylindrical rod, and $\mathbf{B}$, at $n=1$, in the limit in which the ellipsoid becomes a flat pancake~\cite{abrikosov}.}.

\section{Magnetization of two dimensional fermi sea}\label{2D}

In metals, at low temperatures, $f(B)$ can have a negative second derivative, implying the absence of stable homogeneous states. Since the average of $B$ is conserved, the system will phase separate into regions of locally different $B$, a phenomenon known as Condon-Domain formation~\cite{DS,abrikosov}. As we will see these instabilities are not typical at low temperatures for the fractionalized fermi seas of our interest, in spite of them displaying an analogue of dHvA oscillations. Instabilities of this sort will be present however over a region of finite temperatures as described in section~\ref{intermT}.

\subsection{Two dimensional metals}

\begin{figure}[t]
	\begin{center}
		\includegraphics[width=2.7in]{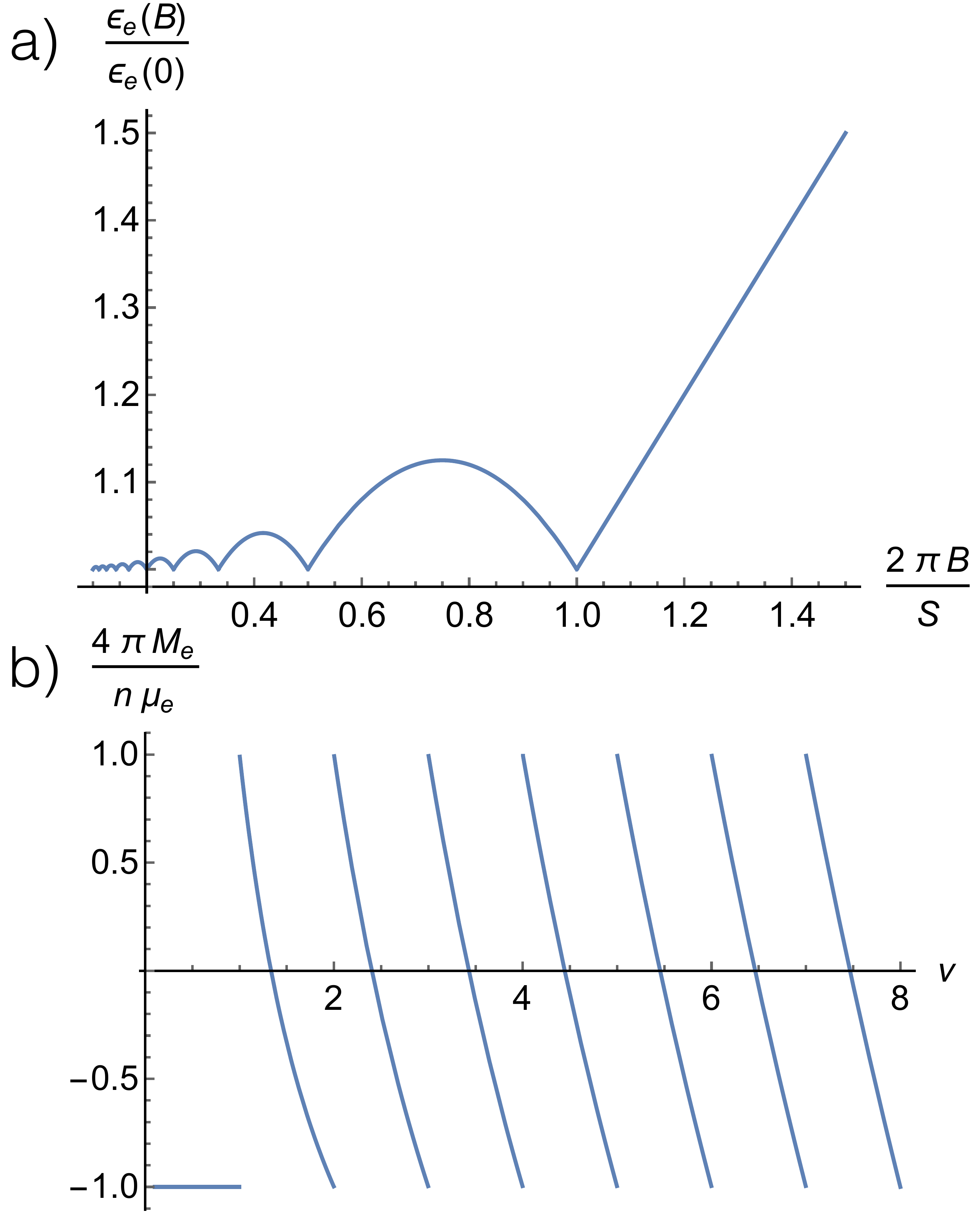}
	\end{center}
	\caption{(Color online) Electronic contribution to (a) energy and (b) magnetization for a two-dimensional electron gas at zero temperature.}
	\label{electron}
\end{figure}

Let us briefly recapitulate the magnetostatics of metals with a single fermi surface in a magnetic field $B$. At zero temperature their energy density is:

\be
u (n,B)= n \epsilon_e(n,B)+\frac{\chi}{2} B^2,
\ee

\noindent where $n$ is the electron density, $\epsilon_e$ the kinetic energy per electron, and $\chi$ the magnetic susceptibility of all background gapped matter including the magnetic energy of vacuum\footnote{Note that in 2D the vacuum contribution contains a prefactor of the film thickness $d$.}. For simplicity we consider spinless electrons. If fermion-fermion interactions can be ignored, and assuming a parabolic dispersion, we have ($\hbar=e=c=1$):

 \be\label{Ee}
 \epsilon_e(n,B) &=& \epsilon_e(n,0)\left(1+\frac{\nu_f(1-\nu_f)}{\nu^2} \right), \\
   \epsilon_e(n,0) &=& \frac{\pi n}{m_e},
 \ee
 
\noindent where $\nu=N/N_\phi=n \Phi_0/B=2 \pi n/B=S/(2 \pi B)$ is the filling of the Landau level spectrum and we write as $\nu=\nu_i+\nu_f$, where $\nu_{i,f}$ are the integer and fractional parts of $\nu$, and $S=\pi k_F^2$ is the area of the Fermi surface. $\epsilon_e$ is depicted in Fig.~\ref{electron}(a). Therefore, the electronic contribution to the magnetization is:

\be
4 \pi M_e \equiv -n \frac{\partial  \epsilon_e}{\partial B}=2\mu_e n \frac{\nu_i(1+\nu_i)}{\nu}-\mu_e n(1+2 \nu_i),
\ee

\noindent where $\mu_e=1/2 m_e$ is the Bohr magneton. The jump in the magnetization, which can be associated with the amplitude of the magnetic oscillations at zero temperature, is given by:

\begin{equation}\label{xx}
\begin{split}
4\pi \delta M_e \equiv &  4\pi  M_e (\nu \rightarrow (n+1)^+)-4\pi  M_e (\nu \rightarrow (n+1)^-) \\
 = &2 \mu_e n. \\
\end{split}
\end{equation}

\noindent Notice that the second derivative of the energy is:
 
 \be
 \frac{\partial^2\epsilon_e(n,B)}{\partial B^2}=-\frac{2 \epsilon_e(n,0)}{(n\varphi_0)^2} \nu_i(1+\nu_i), \ \nu\notin {\mathbb Z}.
 \ee
 
\noindent This implies that for:
  \be
 \nu_i(1+\nu_i)> 2 \pi m_e \chi = \frac{\chi}{12 \chi_e},
 \ee
 
\noindent the metal will be thermodynamically unstable  except at the discrete values of $B$  for which the Landau levels are completely filled $\nu\in {\mathbb Z}$, since the energy curve has a infinite second derivative and is locally concave around these singular points as depicted in Fig.~\ref{electron}. In the above expression, $\chi_e\equiv 1/(24\pi m_e)$ is the Landau diamagnetic susceptibility of a spinless two-dimensional metal with parabolic dispersion.

\subsection{Two dimensional neutral Fermi sea at $T=0$}\label{2DTzero}

\begin{figure*}
\begin{center}
\includegraphics[width=7.0in]{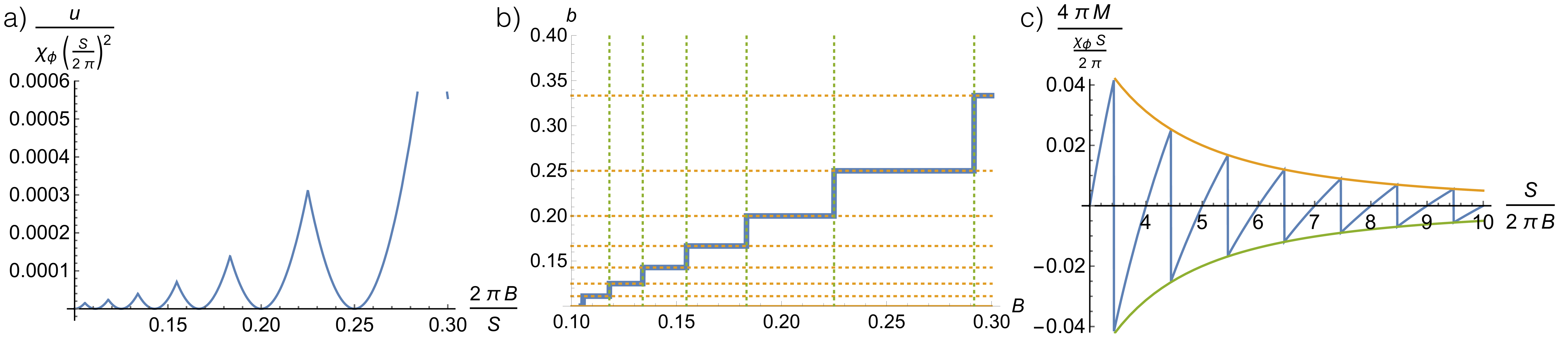}
\end{center}
\caption{(Color online) (a) Contribution of the neutral fermion and the gapped boson in fractionalized neutral fermi sea to energy. (b) Emergent magnetic field as a function of phyisical magnetic field (horizontal and vertical dashed lines indicate values obtained from Eq.~\eqref{bp} and Eq.~\eqref{Bp} respectively.). (c) Contribution to net magnetization (orange and green envelopes correspond to the amplitude obtained from Eq~\eqref{deltaM}). All curves are at zero temperature and in the two-dimensional case.
\label{spinon2D}
}
\end{figure*}

Since the neutral Fermi sea is a uniform state, at a mean field level, we can consider the internal magnetic field, $b=\nabla \times a$, to be uniform. Unlike the physical magnetic field $B$, however, the average value of the internal field $b$ is not an independent thermodynamic quantity\footnote{Strictly speaking $b$ is a conserved quantity at low energies. This is because monopole fluctuations of $b$ are irrelevant in the infrared~\cite{monopoles1,monopoles2}. However these fluctuations will always be present and lead to relaxation of global value of $b$. We will ignore the more subtle issue that this relaxation process might be slow due to irrelevance of monopole fluctuations and assume that on sufficiently long time scales the system relaxes the spatial average of $b$ to the global minimum of the free energy.}. At small $B$, the mean field energy of the system is:

\be\label{u1}
u=n \epsilon_\psi(n,b)+\frac{\chi_\varphi}{2}(b-B)^2+\frac{\chi}{2}B^2.
\ee

\noindent Here $\epsilon_\psi$ is the kinetic energy per particle of the neutral fermions, $n$ is their density, $b$ is the internal magnetic field, $\chi_\varphi$ is the magnetic susceptibility of the gapped bosons, and $\chi$ is the susceptibility of all background trivial gapped matter. The form above follows simply from adding the energy of the boson, $\frac{\chi_\varphi}{2}(b-B)^2$, which has a quadratic form at small fields because the boson is gapped, and the kinetic energy of the neutral fermions. The essential point is that since $b$ is a dynamical degree of freedom it will be self-consistently adjusted to minimize the energy, unlike $B$ which is a fixed thermodynamic parameter. Therefore, operationally, the task is to find the minimum of the energy in Eq.~\eqref{u1} with respect to $b$ at any given $B$.

Let us define the neutral fermion filling factor with respect to $b$ as $\nu_\psi=2\pi n/b=S/(2\pi b)$, and similarly denote $\nu_{\psi i,\psi f}$ the integer and fractional parts of $\nu_\psi$. Notice that under the approximation of a parabolic dispersion for the neutral fermions, $\epsilon_\psi$ has the same form as $\epsilon_e$ given in Eq.~\eqref{Ee} except that $m_e\rightarrow m_\psi$ and $\nu \rightarrow \nu_\psi$. At small enough fields the following condition is satisfied:

 \be
\nu_{\psi i}(1+\nu_{\psi i})> 2 \pi m_\psi \chi_\varphi = \frac{\chi_\varphi}{12 \chi_\psi},
 \ee

\noindent and when this happens, the local minima of Eq.~\eqref{u1} are always achieved at integer filings of the Landau levels of $\psi$. This remarkable tendency, first identified in Motrunich's work~\cite{Motrunich}, can be traced back to the existence of cusps in the energy per particle at integer filling as a function of particle number ({\it i.e} the cyclotron gaps, see Fig.~\ref{spinon2D}(a)). It tells us that the internal magnetic field, $b$, will remain locked at a constant value for a finite range of $B$ so that the neutral fermions fill an integer number of Landau levels within such range, as depicted in Fig.~\ref{spinon2D}(b). Interestingly, any integer filling at a given external $B$, describes a metastable state because it is a local minima of the free energy. The ground state is selected from these metastable states as the one having the absolute lowest energy. These different metastable states can thus be labeled by an integer $p$:

\be\label{bp}
b_p=2\pi n/p, \ p \in {\mathbb Z},
\ee

\noindent and their energy is:

\be
u_p (n,B)=n \epsilon_\psi(n,0)+\frac{\chi_\varphi}{2}(b_p-B)^2+\frac{\chi}{2} B^2,
\ee

\noindent where we have used the special property that the kinetic energy of any number of completely filled Landau levels is identical to the kinetic energy at zero field for Galilean fermions. The critical fields at which the ground state of the system transitions from filling $\nu_\psi=p+1$ to $\nu_\psi=p$ is given by:
\be\label{Bp}
B_p=\pi n \left(\frac{1}{p}+\frac{1}{p+1}\right).
\ee

\noindent For $B_{p+1}\leq B\leq B_{p}$ the fermion filling factor is $\nu_\psi=p+1$. For this range of magnetic fields, the contribution to the magnetization from the neutral fermi sea and the gapped boson is locally a decreasing function of the physical magnetic field:

\be
4 \pi M=-\chi_\varphi  \left(B-\frac{2\pi n}{p+1}\right).
\ee

\noindent The behavior of this magnetization is depicted in Fig.~\ref{spinon2D}(c). Remarkably, the period of the oscillations is controlled by the {\it bare} magnetic field. In other words the oscillations are periodic in $1/B$ with a period given by:

\be\label{periodT=0}
\Delta \left(\frac{1}{B}\right)=\frac{1}{2\pi n}=\frac{2\pi}{S}, \ B\ll S, \ T=0,
\ee 

\noindent where $S$ is the area of the Fermi surface. Let us thus denote by $\nu=2\pi n/B$. Therefore, to find the integer filling describing the ground state, $p(\nu)$, at a given magnetic field $B$, we need to solve the following inequality:

\be
\frac{2 p (p+1)}{2 p+1}\leq \nu \leq \frac{2 (p+1) (p+2)}{2 p+3}, \ p\in {\mathbb Z}, \ \nu\in {\mathbb R}.
\ee

\noindent For $p\gg 1$ the inequality reduces to: $p+\mathcal{O}(1/p) \leq \nu-1/2\leq p+1+\mathcal{O}(1/p)$, and thus can be solved by $p\approx[\nu-1/2]$. The size of the jump of the magnetization at the transition from $\nu_\psi=p+1$ to $\nu_\psi=p$ is therefore:

\begin{equation}\label{deltaM}
\begin{split}
4\pi \delta M\equiv &4 \pi M(B\rightarrow B_p^-)-4 \pi M(B\rightarrow B_p^+)=\frac{2 \pi n \chi_\varphi}{ p(p+1)} \\
 &\approx 2 \pi n \chi_\varphi \left(\frac{B}{2 \pi n}\right)^2, \ p\gg1.\\
\end{split}
\end{equation}

Notice that the neutral fermi sea displays generically a concave energy as a function of $B$ except at discrete points, as illustrated in Fig.~\ref{spinon2D}(a). Therefore, the spinon fermi surface does not suffer from the generic low-temperature thermodynamic instabilities that we encountered in a metal (compare Figs.~\ref{electron}(a) and~\ref{spinon2D}(a)). The physical picture for why this happens is that the internal magnetic field can adjust itself until the neutral fermions completely fill an integer number of Landau levels, and when this happens the system enjoys the stability of being able to form a uniform fully gapped state\footnote{This state is technically a chiral spin liquid. It is also easy to argue that this state is stable beyond the simple mean field picture we describe in the main text, because the gauge field fluctuations will be gapped at long wavelengths due to the Chern-Simons term induced by the neutral fermions forming an integer quantum Hall state.}.

Since the magnetization of the neutral fermions is locally a decreasing function of $B$ (Fig.~\ref{spinon2D}(c)), one could say that the neutral fermion system tends to be {\it differentially diamagnetic}, unlike electrons which tend to be {\it differentially paramagnetic} at low temperatures. This distinction will be washed out at higher temperatures. Notice, however, that the magnetization of the neutral fermion is not a strictly decreasing function of $B$ because the curve is piece-wise discontinuous (Fig.~\ref{spinon2D}(c)). 

\subsection{Two-dimensional neutral Fermi sea at finite $T$}\label{finiteT}

In this section, we consider the magnetic response of the system at a finite temperature when the neutral fermion cyclotron spacing is much smaller than the temperature broadening but the temperature is much smaller than the fermi energy of the neutral fermion and any other scale associated with the gap of the boson, specifically $n/m_\psi \gg T\gg b/m_\psi$. To leading order in small $b$ and $B$ the free energy of the system is:

\be\label{f1}
f=f_0(B)+\frac{\chi_\psi}{2} b^2+\frac{\chi_\varphi}{2} (B-b)^2,
\ee

\noindent where $\chi_\psi$ characterizes the diamagnetic coefficient of the neutral fermi sea that expresses the energy cost for $b$ to be non-zero ({\it i.e.} the analogue of the Landau diamagnetic susceptibility of a metal) and it is given by (spinless fermions): $\chi_\psi=1/(24\pi m_\psi)$, but we will keep it as an unspecified coefficient for generality, and $f_0$ includes the contribution of the background trivial gapped matter and the magnetic energy of vacuum.

\be\label{bB}
b_B=\frac{\chi_\varphi}{\chi_\psi+\chi_\varphi} B \equiv \alpha B.
\ee

\be
B-b_B=\frac{\chi_\psi}{\chi_\psi+\chi_\varphi} B=(1-\alpha) B.
\ee

\begin{figure}
\begin{center}
\includegraphics[width=0.9\columnwidth]{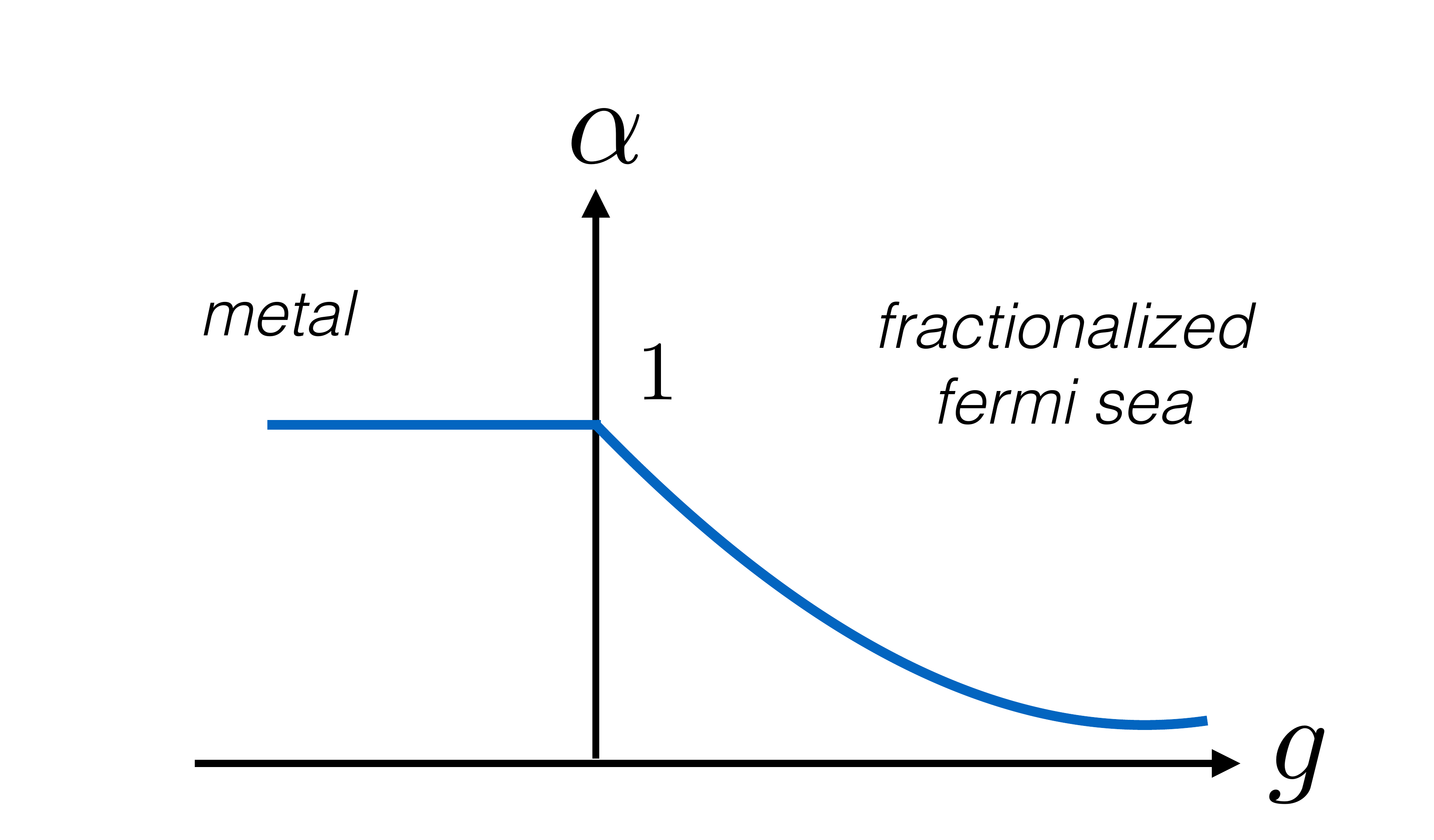}
\end{center}
\caption{(Color online) Qualitative behavior of the parameter $\alpha$ that controls the effective magnetic field experienced by the neutral fermions in response to a physical magnetic field across the transition from metal to fractionalized neutral fermi surface, driven by a parameter $g$ which controls whether the boson is gapped ($g>0$) or condensed ($g<0$).}
\label{alpha}
\end{figure}

\noindent Therefore, for $|\chi_\psi|\ll |\chi_\varphi|$ the internal field experienced by the fermions tends to track the external field ($\alpha\rightarrow 1$), while in the opposite limit the neutral fermions experience just a small fraction of the field while the bosons experience almost the full external field ($\alpha\rightarrow 0$). The latter is typically what one expects if the bosons are deep in an insulating phase, as their energy would hardly change in the presence of an effective field, $B-b$, implying that $\chi_\varphi \rightarrow 0$ as we move deeper into the insulator. The opposite behavior, $\chi_\varphi \rightarrow \infty$, is what one expects if the boson is condensed, developing a Meissner effect for the effective field, $B-b$. Across a continuous phase transition between the insulator and the metal~\cite{FG2004,TS2008}, assuming the transition is of the kind that occurs at fixed boson density, one expects that $\chi_\varphi$ diverges as one approaches the metal from the insulating side because in the metal the boson has condensed. This implies that in the phase transition between an ordinary metal and the fractionalized neutral fermi sea, the coefficient $\alpha$ that relates the effective field experienced by the neutral fermions, $b$, to the physical magnetic field, $B$, starts continuously decreasing from $1$ at the critical point (see Fig.~\ref{alpha})\footnote{It is worth bearing in mind that the value of this coefficient in the fractionalized phase is not universal and can be even larger than $1$ under special circumstances, as found by Motrunich due to the proximity to the uniform flux phase of the spinon fermi sea that he considered~\cite{Motrunich}.}.  

\noindent At small fields and at temperatures higher than the cyclotron energy, the contribution to the physical magnetization of the neutral fermi sea depends linearly on the external field:

\be\label{Mlinear}
4\pi M=-\left(\frac{1}{\chi_\varphi^{-1}+\chi_\psi^{-1}}\right) B\equiv -\chi_0 B.
\ee

\noindent The physical content of this expression is that whichever of the two components (i.e. the fermions or the bosons) is less responsive in changing its energy in the presence of an effective magnetic field, will dominate the magnetization response to external fields. Notice that this is opposite to the case of ``stacking" two physical systems for which the respective $\chi$'s add up and hence the more responsive component dominates. This result is an example of the well-known Ioffe-Larkin rule for addition of electromagnetic response functions for partons~\cite{IL}.

\subsection{Magnetic oscillations of two-dimensional neutral fermi sea at finite $T$}\label{oscT}

The oscillatory part of the free energy of the neutral fermi sea can be estimated at a mean field level by using the classic result from two-dimensional metals~\cite{DS}:

\be\label{fosc}
f_{\rm osc}=\frac{\chi_{\rm osc} b^2}{2} \sum_{k=1}^\infty \frac{(-1)^k}{k^2} \frac{\frac{k\tau S}{2\pi b} }{\sinh(\frac{k\tau S}{2\pi b} )} \cos\left(\frac{k S}{b} \right).
\ee

\noindent Here $S$ is the area of the neutral Fermi surface, $S=\pi k_F^2=4 \pi^2 n$, $\tau=2\pi^2 T/\epsilon_F$, $\epsilon_F=k_F^2/(2m_\psi)$ and $\chi_{\rm osc}$ is a parameter controlling the amplitude of the oscillations, which for spinless parabolic fermions is $\chi_{\rm osc}=1/(2 \pi^3 m_\psi)$. This form also holds for arbitrary dispersions within an effective mass approximation around the Fermi surface, with corrections from non-parabolicity appearing as higher powers in $(b/S)$. The task is to minimize the sum of the energies from Eqs.~\eqref{f1} and~\eqref{fosc} as a function of $b$. \noindent At low temperatures, namely when the temperature is lower than the effective cyclotron energy of the neutral fermions ($T\ll \omega_\psi \equiv b/m_\psi$), and at low $b$ ($b\ll S$), the minima of the free energy will be dominated by the oscillatory part, because this term will dominate the derivative of the energy due to the divergence of its oscillation frequency as $1/b$ for $b\rightarrow 0$. This allows to recover the result found in Section~\ref{2DTzero}, that the minima will be pinned at a fixed value of $b$ within a range of $B$. These correspond to the minima of $f_{\rm osc}$ which are given by:

\be
b_p=\frac{S}{2\pi p}=\frac{2 \pi n}{p}, \ p\in {\mathbb Z}.
\ee

\noindent One thus recovers the result of Eq.~\eqref{bp}. Therefore the free energy of the $p-$th metastable state can be approximated at low temperature as:

\begin{equation}\label{fprime}
\begin{split}
f_p\approx & \tilde{f}_0+\frac{\chi_\psi+\chi_\varphi}{2} (b_p-b_B)^2+\frac{\chi_{osc} }{2} b_p^2 \left(\sum_{k=1}^\infty \frac{(-1)^k}{k^2} \right), \\
= &  \tilde{f}_0+\frac{\chi_\psi+\chi_\varphi}{2} (b_p-b_B)^2-\frac{ \pi^2\chi_{osc} }{24} b_p^2,
\end{split}
\end{equation}

\noindent where $b_B$ was defined in Eq.~\eqref{bB} and $\tilde{f}_0=f_0+\chi_0 B^2/2$ includes the linear in $B$ contribution to magnetization from Eq.~\eqref{Mlinear}. Following a similar line of reasoning as in Section~\ref{2DTzero}, one finds the following behavior for the amplitude of the oscillatory part of the magnetization at low temperatures:

\begin{equation}\label{Mosc}
\begin{split}
4\pi \delta M_{osc} \approx & \frac{\chi _{\varphi }^3}{(\chi _{\varphi }+\chi _{\psi }-\pi^2 \chi_{osc}/12)^2} \frac{S}{2\pi} \left(\frac{2\pi B}{S}\right)^2,\\
= &  2\pi n \chi _{\varphi } \left(\frac{B}{2\pi n}\right)^2, \ B\ll S, \ T\ll \omega_\psi.
\end{split}
\end{equation}

\noindent At low temperatures ($T\ll \omega_\psi$) the period of the oscillations can be estimated to be:

\begin{equation}\label{DeltaBlow}
\begin{split}
\Delta \left(\frac{1}{B}\right)\approx &\left(\frac{\chi_\varphi}{\chi_\psi+\chi_\varphi-\pi^2 \chi_{osc}/12}\right) \frac{2\pi}{S},\\
= & \frac{2\pi}{S}, \ B\ll S
\end{split}
\end{equation}

\noindent where in the second line of the two preceding equations we have used the relation $\chi_\psi=\pi^2 \chi_{osc}/12$, valid for parabolic fermions, and we have recovered the results of Eq.~\eqref{deltaM} and Eq.~\eqref{periodT=0}.  Now, at higher temperature ($T\gtrsim \omega_\psi \equiv b/m_\psi$) the oscillatory contribution to the free energy is negligible compared to the smooth part from Eq.~\eqref{f1}. In this case, the minimum of the free energy will be achieved when $b=b_B=\alpha B$, and generically there is a single solution and no multiple metastable states. In this regime the free energy of the equilibrium state can therefore be approximated as:

\begin{equation}\label{}
\begin{split}
f\approx & \tilde{f}_0+f_{\rm osc}(b\rightarrow b_B). \\
\end{split}
\end{equation}

\noindent Therefore the functional form of the magnetic oscillations of the neutral fermi sea at higher temperature are essentially the same as that of a metal but in an effective field given by $b_B$. Therefore the leading contribution to the oscillatory magnetization at small $b$ ($b_B\ll S$) is:

\begin{equation}\label{Mosc2}
\begin{split}
4 \pi  M_{\rm osc} \approx & \frac{\chi_{\rm osc} \alpha S}{2} \sum_{k=1}^\infty \frac{(-1)^{k+1}}{k} \frac{\frac{k S \tau }{2 \pi  b_B}}{ \sinh \left(\frac{k S \tau }{2 \pi  b_B}\right)} \sin \left(\frac{k S}{b_B}\right) \\
&   \ b_B\ll S, \ \epsilon_F \gg T\gtrsim \omega_\psi.
\end{split}
\end{equation}

\noindent An interesting feature of the oscillations in two dimensions is that the period of the oscillations has different behavior at low temperatures ($T\ll \omega_\psi$) than at high temperatures ($\epsilon_F\gg T\gtrsim \omega_\psi$). While at low temperatures we found a period given in Eq.~\eqref{DeltaBlow}, at higher temperatures we have

\be\label{}
\Delta \left(\frac{1}{B}\right)=\frac{2\pi \alpha}{S_\perp}, \ B\ll S, \ T\gtrsim \omega_\psi.
\ee

\noindent On the other hand, at high temperature, the amplitude of the oscillations is therefore given by,

\begin{equation}\label{}
\begin{split}
4 \pi \delta M_{\rm osc} \approx & \chi_{\rm osc} \alpha S \sum_{k=1}^\infty \frac{(-1)^{k+1}}{k} \frac{\frac{k S \tau }{2 \pi  b_B}}{ \sinh \left(\frac{k S \tau }{2 \pi  b_B}\right)} \\
&   \ b_B\ll S, \ \epsilon_F \gg T\gtrsim \omega_\psi.
\end{split}
\end{equation}

\noindent Notice that if we were to naively extrapolate the amplitude of the oscillations to the low temperature regime, we would obtain the following amplitude:
$4 \pi \delta M^{\rm naive}_{\rm osc} \approx  \chi_{\rm osc} \alpha S \log (2), \ B\ll S, \  T\ll b_B/m_\psi$, which would overestimate the correct result given in Eq.~\eqref{Mosc}. Similarly if we extrapolate the low temperature result to high temperatures we would overestimate the result of Eq.~\eqref{Mosc2}. This motivates a simple functional form that interpolates between the two regimes and that captures reasonably well the crossover region:

\begin{equation}\label{interp}
\begin{split}
& \frac{1}{4 \pi \delta M_{\rm osc}} \approx   \\
&  \frac{1}{4 \pi \delta M_{\rm osc}(T\ll \omega_\psi)}+\frac{1}{4 \pi \delta M_{\rm osc}(T\gtrsim \omega_\psi)},
\end{split}
\end{equation}

\noindent where the formulas for the right hand side are given in Eqs.~\eqref{Mosc} and~\eqref{Mosc2} respectively. A simple approximation is to keep the leading harmonic in the oscillatory part, as is done in the celebrated Lifshitz-Kosevich formula for metals, to obtain
\be\label{deltaMosc3D}
\frac{1}{4 \pi \delta M_{\rm osc}}\approx \frac{1}{\chi _{\varphi } \frac{S}{2\pi} \left(\frac{2 \pi B}{S}\right)^2}+\frac{1}{\chi_{\rm osc} \alpha S \frac{\frac{S \tau }{2 \pi  b_B}}{ \sinh \left(\frac{S \tau }{2 \pi  b_B}\right)}}.
\ee

\section{Magnetization of three dimensional fermi sea}\label{3DfiniteT}

\begin{figure}[t]
	\begin{center}
		\includegraphics[width=2.9in]{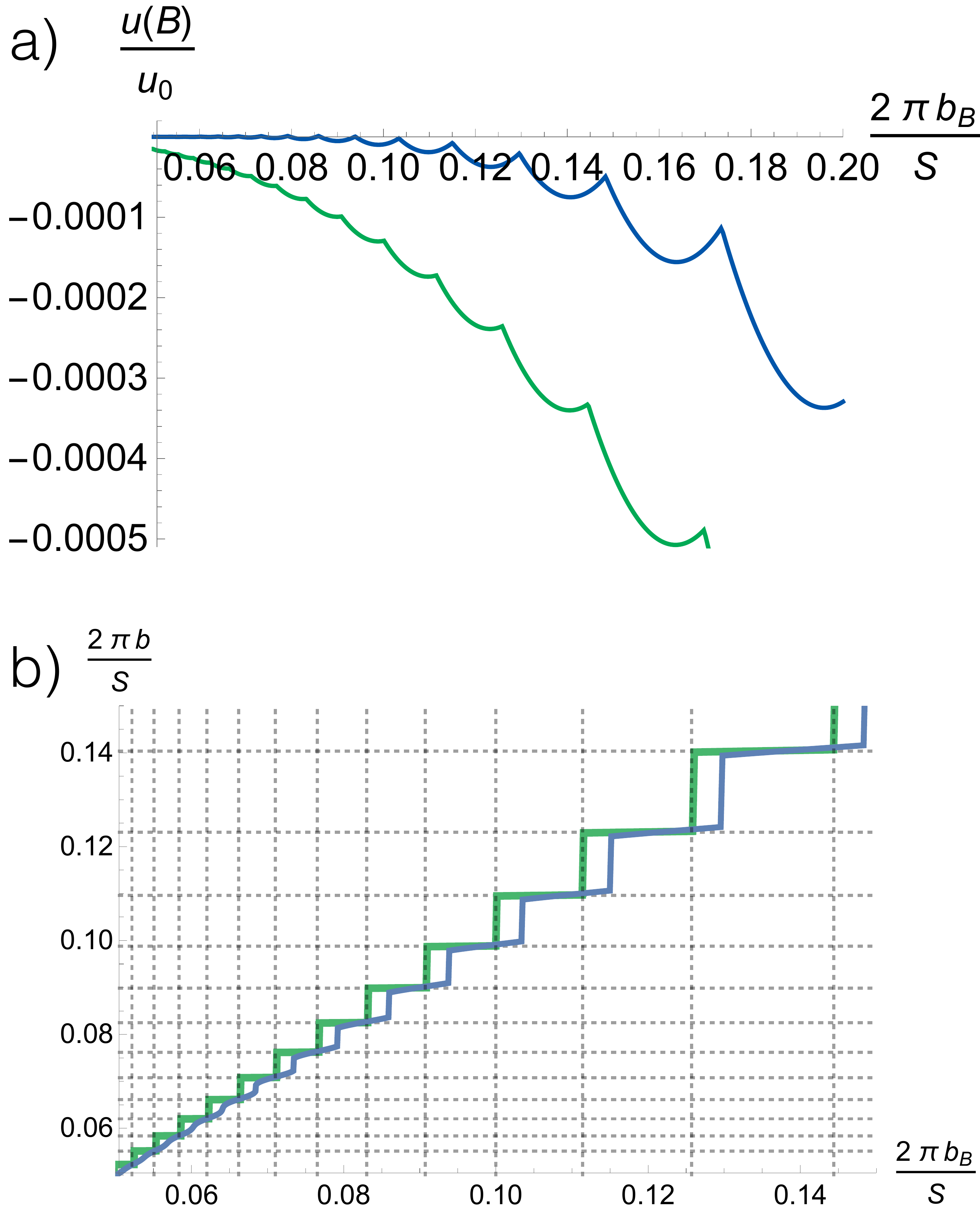}
	\end{center}
	\caption{(Color online) (a) Energy of three-dimensional fractionalized neutral fermi sea as a function of magnetic field (the scale is $u_0=(\chi_\varphi+\chi_\psi)(S_\perp/2\pi)^2$) (b) Internal field $b$ as function of the rescaled external field $b_B=\alpha B$. Dashed horizontal and vertical lines correspond to the values expected from Eqs.~\eqref{bp3D} and~\eqref{Bp3D} respectively. Both curves were computed from direct numerical minimization of the model described in Section~\ref{intermT}, and parameters were chosen as $\chi_{\rm osc}/(\chi_\varphi+\chi_\psi)=0.1$ and effective temperatures $\tau=0.1$ (green) and $\tau=0.5$ (blue). In the energy, the smooth background term associated with Eq.~\eqref{Mlinear} is not included.}
	\label{spinon3Da}
\end{figure}

There are several important qualitative differences between three and two-dimensions that we now wish to emphasize. First of all, gauge-field fluctuations are weaker in three-dimensions and hence one expects the kind of mean field treatment that we have outlined to be more reliable. On the other hand, in three dimensions the spectrum of the fermions in a magnetic field is that of Landau bands that disperse along the direction of the magnetic field instead of Landau levels, which leads to important quantitative differences with respect to the two-dimensional case. In spite of this difference, the analysis in three-dimensions parallels in several ways that in two dimensions. For example, we will find in three dimensions the same tendency for the internal magnetic field to remain pinned at certain discrete values at ultra-low temperatures, in spite of the presence of Landau bands rather than Landau levels. Hence, the ultra-low temperature quantum oscillations will become a series of phase transitions between different metastable states, similar to two-dimensions. Similarly, the high temperature behavior of the oscillations will again resemble those of an ordinary metal experiencing an effective magnetic field $b_B=\alpha B$; an important qualitative difference with two-dimensions is that the period of the oscillations will not change with temperature.

\subsection{Three dimensional neutral Fermi sea at finite $T$}

Consider a three dimensional fractionalized Fermi sea. For generality, we assume an anisotropic dispersion with effective mass $m_z$ along the $z-$axis and $m_\perp$ in the $xy-$plane, and take the magnetic field to act along the $z-$direction. Now at temperatures that are higher than the transverse cyclotron energy ($\epsilon_F \gg T\gg b/m_\perp$), the free energy has the same form as in Eq.~\eqref{f1}

\be\label{f03D}
f=f_0+\frac{\chi_\psi}{2} b^2+\frac{\chi_\varphi}{2} (B-b)^2,
\ee

\noindent except that the coefficients are different (for spinless parabolic fermions $\chi_\psi=\nu_F/(12 m_\perp^2)=n/(8  m_\perp^2 \epsilon_F)$, $\nu_F=3n/(2\epsilon_F)$, $n=((2\epsilon_F)^{3} m_z m_\perp^2)^{1/2}/(6 \pi^2)$). Once again, the minimum of the energy is achieved at the same field and the contribution to the magnetization from the partons has the same form,

\begin{equation}\label{}
\begin{split}
&b_B=\frac{\chi_\varphi}{\chi_\psi+\chi_\varphi} B \equiv \alpha B,\\
&4\pi M=-\left(\frac{1}{\chi_\varphi^{-1}+\chi_\psi^{-1}}\right) B\equiv -\chi_0 B.
\end{split}
\end{equation}

\noindent Now at lower temperatures, the oscillatory part of the free energy becomes important and we will have an extra contribution of the form~\cite{DS,abrikosov},

\begin{equation}\label{fosc3d}
\begin{split}
f_{\rm osc}=&\frac{\chi_{\rm osc} b^2}{2} \left(\frac{2\pi b}{S_\perp}\right)^{1/2} \times  \\
 &\sum_{k=1}^\infty \frac{(-1)^k}{k^{5/2}}\frac{\frac{k\tau S_\perp}{2\pi b} }{\sinh(\frac{k\tau S_\perp}{2\pi b} )} \cos\left(\frac{k S_\perp}{b} \mp \frac{\pi}{4}\right),
\end{split}
\end{equation}

\noindent where $\chi_{\rm osc}=(m_z \epsilon_F)^{1/2}/(4\pi^2m_\perp)$, $S_\perp$ is the area of the cross-section of the Fermi surface perpendicular to the magnetic field, and the $-$($+$) signature is for an electron-like (hole-like) dispersion in the direction of the field, namely for $m_z>0$ ($m_z<0$). Much like in metals, to derive this expression one does not need to assume that the dispersion of the fermions is exactly parabolic. More generally, the dominant contribution to the field dependence of the free energy at small fields ($b\ll S_\perp$) arises from the cross sections of extremal area of the fermi surface that are orthogonal to the applied magnetic field. One assumption is that there are no non-trivial Berry surface terms, but otherwise the formula can be applied if $m_z$ is understood to arise from the curvature of the dispersion near a given extremal cross section and $m_\perp$ is understood as the cyclotron mass for states near such a cross section. The mathematical details of the derivation in this more general case parallel identically those for a conventional metal provided in Ref.~\cite{abrikosov}. 

Similar to the situation in tw dimensions, at low temperatures ($T\ll b/m_\perp$) and small fields ($b\ll S_\perp$) the derivative of the free energy will be dominated by the oscillatory part, due to the diverging oscillation frequency as $1/b$. Therefore the minima of the free energy will be controlled by the oscillatory part, which is independent of $B$. Therefore we encounter again a multiplicity of metastable states associated with the different local minima of the oscillatory part of the free energy, which are approximately given by:

\begin{figure}[t]
	\begin{center}
		\includegraphics[width=2.9in]{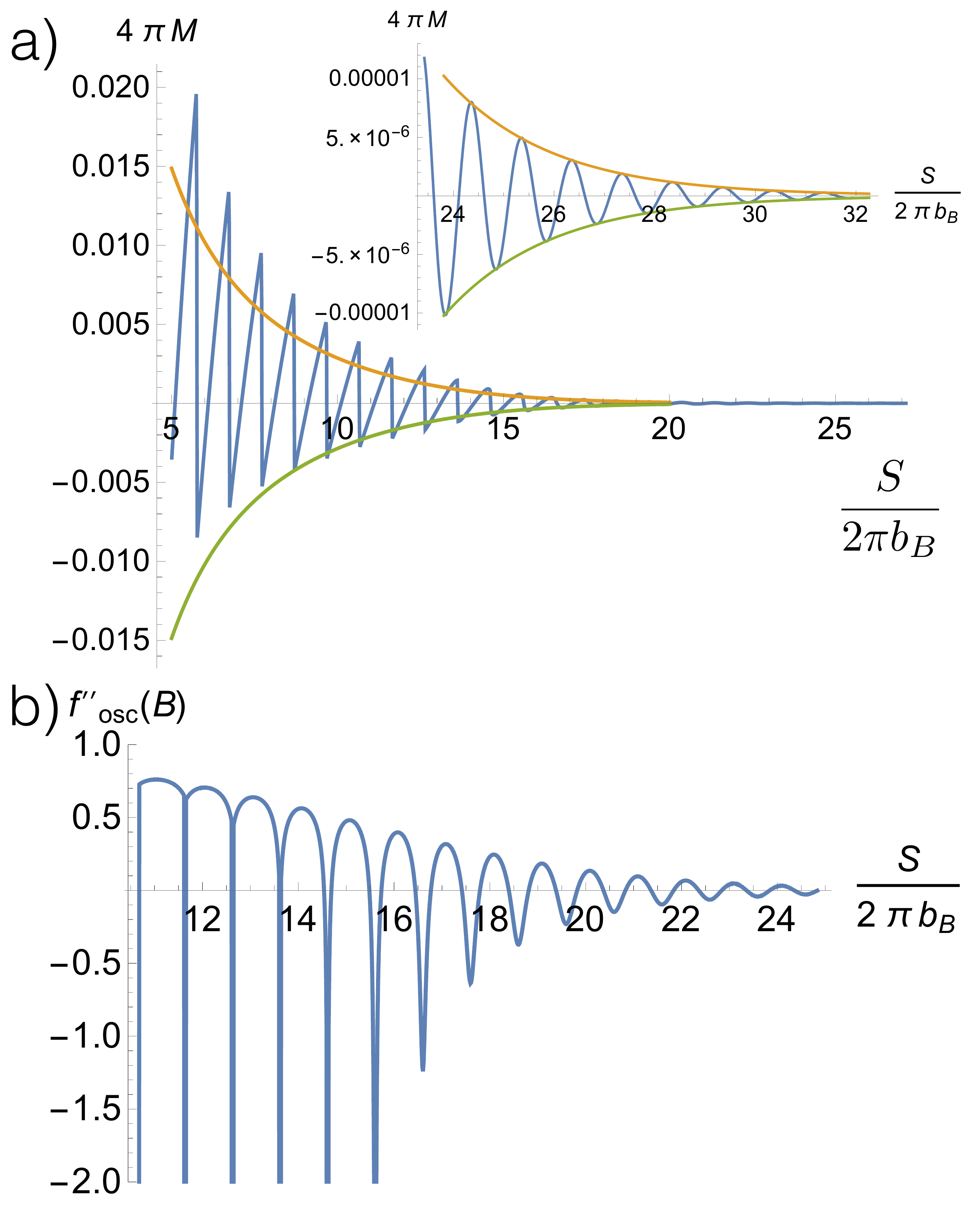}
	\end{center}
	\caption{(Color online) (a) Oscillatory magnetization of a three-dimensional fractionalized neutral fermi sea as a function of magnetic field (the scale is $4 \pi M_0=\chi_\varphi(S_\perp/2\pi)$). The green and yellow envelopes correspond to the amplitude expected from the simplified formula of Eq.~\eqref{deltaMosc3D}. The inset is a zoom-in to the small field region illustrating the good agreement with the formula. (b) The second derivative of the oscillatory component of the free energy with respect to the effective field $b_B=\alpha B$. The curve shows clearly how the delta function-like spikes in the second derivative, occurring at the discontinuous jumps of the magnetization, transform into the regions where the second derivative is smooth but negative where instabilities might appear (see discussion in Section~\ref{intermT}). Both curves were computed from direct numerical minimization of model described in Section~\ref{intermT}, and parameters were chosen as $\chi_{\rm osc}/(\chi_\varphi+\chi_\psi)=0.1$ and effective temperature $\tau=0.5$. The smooth background term associated with Eq.~\eqref{Mlinear} is not included.}
	\label{spinon3Db}
\end{figure}

\be\label{bp3D}
b\approx b_p\equiv \frac{S_\perp}{2 \pi  p\pm\pi/4}, \ p\in {\mathbb Z}.
\ee

\noindent This staircase behavior of the internal field is depicted in Fig~\ref{spinon3Da}(b) where we show the explicit numerical solution of a simplified model that includes a single harmonic of the oscillatory free energy described in Sec.~\ref{intermT}. On a given metastable state, the free energy can therefore be approximated as,

\begin{equation}\label{}
\begin{split}
f_p\approx & \tilde{f}_0+\frac{\chi_\psi+\chi_\varphi}{2} (b_p-b_B)^2-  \frac{c \chi_{osc}}{2} b_p^2 \left(\frac{2\pi b_p}{S_\perp}\right)^{1/2}, \\
\end{split}
\end{equation}

\noindent where $c$ is a pure number which corresponds to the absolute value of the minimum of the periodic function $\sum_{k=1}^\infty \frac{(-1)^{k}}{k^{5/2}} \cos\left(\frac{k S_\perp}{b} \mp \frac{\pi}{4}\right)$, which is approximately $c\approx 0.87$.  The critical field at which the system switches from the $(p+1)$ to the $p$ metastable state can be found by solving for $f_p=f_{p+1}$, and is found to be,

\be\label{Bp3D}
b_B(p)\approx \frac{b_p+b_{p+1}}{2}- \frac{c \chi_{\rm osc} \left(\frac{2\pi}{S_\perp}\right)^{1/2}}{\chi_\psi+\chi_\varphi}\frac{b_p^{5/2}-b_{p+1}^{5/2}}{2(b_p-b_{p+1})}.
\ee

\noindent For $b_B(p)<b_B<b_B(p-1)$, the internal magnetic field will remain pinned at the value $b=b_p$, and the contribution to the physical magnetization coming from the boson and the neutral fermion, including their linear contribution described in Eq.~\eqref{Mlinear}, is

\be
4 \pi M_p\approx \chi_\varphi (b_p-B).
\ee

\noindent It is interesting to note that the differential susceptibility of the Fermi sea plus the charged gapped-boson is diamagnetic and coincides with the susceptibility of the charged boson in the bare external field $B$. The physical picture behind this phenomenon in the low temperature regime is that when the system is in the $p$-th metastable state, the internal field and the fermionic energy become pinned and do not change with small changes of the external magnetic field, leaving the energy of the boson as the only changing part in response to the external magnetic field that contributes to the magnetization. The behavior of the magnetization is depicted in Fig~\ref{spinon3Db}(a), obtained from an explicit numerical solution of a simplified model described in Sec.~\ref{intermT}. The amplitude of the magnetization jump between adjacent metastable states (i.e. amplitude of oscillations) can be estimated to be

\be
4 \pi \delta M \approx \chi_\varphi \frac{S_\perp}{2 \pi} \left(\frac{2 \pi b_B}{S_\perp}\right)^2, \ b_B\ll S_\perp, \ T\ll b_B/m_\perp,
\ee

\begin{figure}[t]
	\begin{center}
		\includegraphics[width=3in]{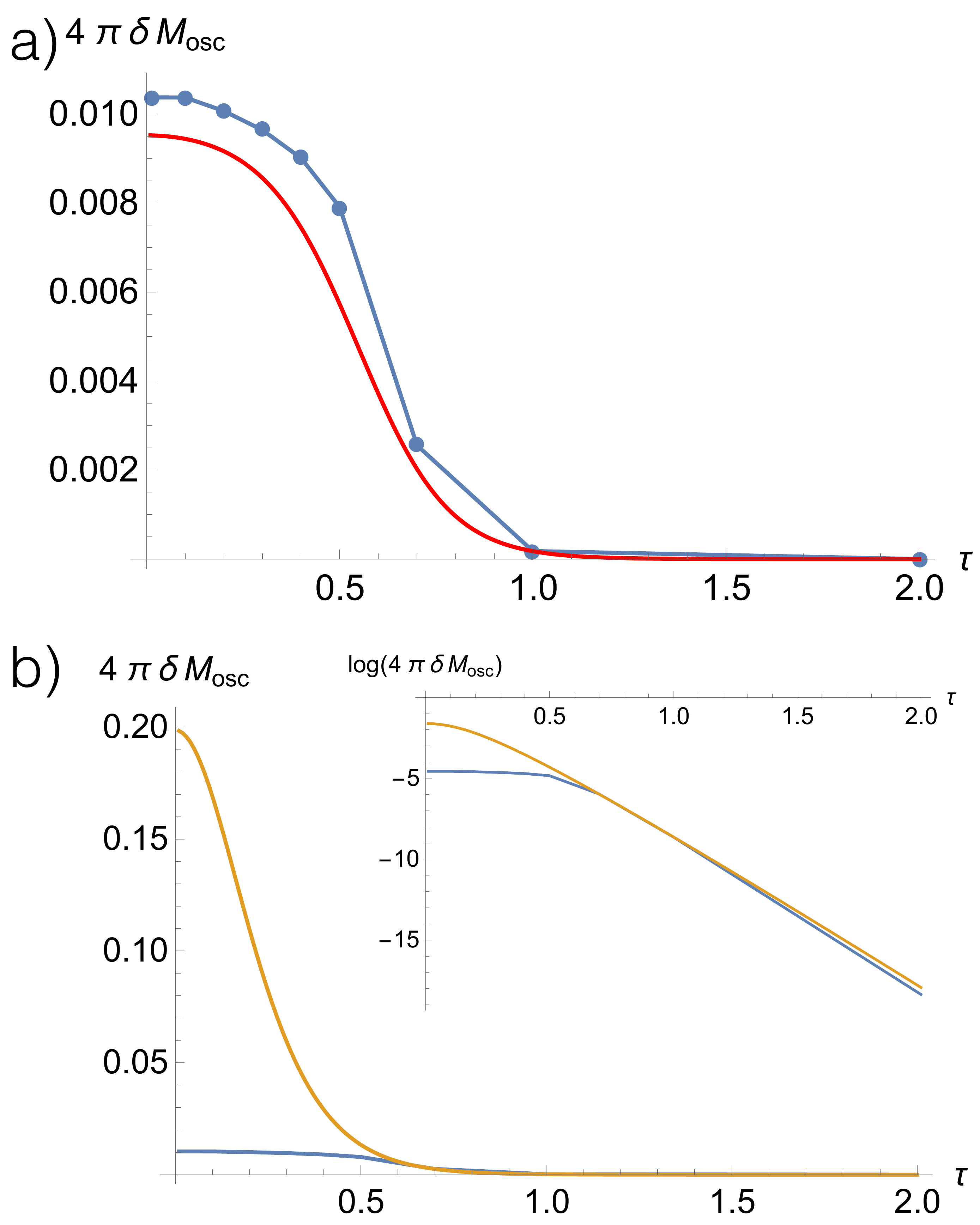}
	\end{center}
	\caption{(Color online) (a) Amplitude of the magnetization oscillations in fractionalized neutral fermi surfaces as a function of dimensionless temperature $\tau$. The blue dots correspond to direct result from numerical study of model described in Section~\eqref{intermT} and the red line is the behavior expected from the Eq.\eqref{deltaMosc3D}. (b) Comparison of the amplitude of the fractionalized neutral fermi surface (blue) from (a) with that expected for a conventional metal (orange) with the same effective mass and subjected to the same effective magnetic field strength (to generate the orange plot we simply use the formula valid for $T\gg b_B/m_\perp$ but plot it over the entire temperature range). Parameters were chosen as $\chi_{\rm osc}/(\chi_\varphi+\chi_\psi)=0.1$ and effective field $2\pi b_B/S_\perp=0.1$.}
	\label{spinon3Damp}
\end{figure}

\noindent which is similar to that found in Eq.~\eqref{Mosc} in the two-dimensional case. Interestingly, and in sharp contrast to the two dimensional case, the period of the oscillations at low temperatures is still governed by the average field $b_B$, the same that as we shall see governs the oscillations at higher temperature. The period of the oscillations can be obtained from Eq.~\eqref{Bp3D}, by computing the range of $1/B$ over which the solution remains in the $p$-th metastable state. Following this, at low temperatures, one finds that the oscillatory portion of the magnetization will be periodic as a function of $1/B$ with a period given by

\be\label{DeltaB}
\Delta \left(\frac{1}{B}\right)=\frac{2\pi \alpha}{S_\perp}, \ b_B\ll S_\perp, \ T\ll b_B/m_\perp,
\ee 
 
\noindent namely the periodicity is that of a metal but in a field reduced by the factor $\alpha =\chi_\varphi/(\chi_\psi+\chi_\varphi)$.
 
Let us now consider the behavior of the oscillations at higher temperatures ($\epsilon_F \gg T\gtrsim b_B/m_\perp$). Here the contribution to the free energy from the oscillatory component is negligible, and therefore the energetics of the internal field is dominated by the smooth part of the free energy given in Eq.~\eqref{f03D}. As a consequence, there is generically a single equilibrium state and not the plethora of metastable states realized at lower temperatures. Similar to the two dimensional case, the free energy of the equilibrium state can therefore be approximated as

\begin{equation}\label{}
\begin{split}
f\approx & \tilde{f}_0+f_{\rm osc}(b\rightarrow b_B). \\
\end{split}
\end{equation}

\noindent Therefore the functional form of the magnetic oscillations of the neutral fermi sea at higher temperatures are essentially the same as that of a metal but in an effective field given by $b_B$. Therefore the leading contribution to the oscillatory magnetization at small $b$ ($b_B\ll S$) is

\begin{equation}\label{Mosc3D}
\begin{split}
4 \pi  M_{\rm osc} \approx & \frac{\chi_{\rm osc} \alpha S_\perp}{2} \left(\frac{2\pi b_B}{S_\perp}\right)^{1/2} \times \\
&\sum_{k=1}^\infty \frac{(-1)^{k+1}}{k^{3/2}} \frac{\frac{k S \tau }{2 \pi  b_B}}{ \sinh \left(\frac{k S \tau }{2 \pi  b_B}\right)} \sin \left(\frac{k S}{b_B} \mp \frac{\pi}{4}\right), \\
&   \ b_B\ll S, \ \epsilon_F \gg T\gtrsim \omega_\psi.
\end{split}
\end{equation}

\noindent As in two-dimensions, we can propose an interpolation between the low and intermediate temperature regimes for the amplitude of magnetic oscillations following the formula~\eqref{interp}. A simple approximation is to keep the leading harmonic in the oscillatory part, as is done for the Lifshitz-Kosevich formula of metals, to obtain

\begin{equation}\label{interp3D}
\begin{split}
& \frac{1}{4 \pi \delta M_{\rm osc}}\approx   \\
&  \frac{1}{\chi_\varphi \frac{S_\perp}{2 \pi} \left(\frac{2 \pi b_B}{S_\perp}\right)^2}+\frac{1}{\chi_{\rm osc} \alpha S_\perp \left(\frac{2\pi b_B}{S_\perp}\right)^{1/2} \frac{\frac{\tau S_\perp  }{2 \pi  b_B}}{ \sinh \left( \frac{\tau S_\perp  }{2 \pi  b_B}\right)}}.
\end{split}
\end{equation}

\noindent In Fig.~\ref{spinon3Damp} we illustrate the behavior of the amplitude obtained from an explicit numerical solution of the model with a free energy with a single harmonic described in section~\ref{intermT}. We see that Eq.~\eqref{interp3D} captures the low and high-temperature behavior accurately and provides a good match for the crossover region as well.

Interestingly, in three dimensions, the period of the oscillations at higher temperatures coincides with that at low temperatures (Eq.~\eqref{DeltaB}), and can be read off from Eq.~\eqref{Mosc3D} to be

\be
\Delta \left(\frac{1}{B}\right)=\frac{2\pi \alpha}{S_\perp}, \ b_B\ll S_\perp.
\ee 

We would like to conclude this section by briefly discussing the impact of disorder. The impact of weak short range impurities in the amplitude of oscillations is qualitatively similar to that of temperature~\cite{abrikosov}. Collisions off impurities will broaden the energy levels of the fermions on a scale of the order of the scattering rate $\sim 1/\tau_{\tn{imp}}$. Therefore, at a mean field level, heuristically the impact of disorder can be captured by replacing the temperature in all of our discussion with $\sim T+1 /\tau_{\tn{imp}}$. More specifically, the impact of impurities can be captured by adding an additional suppression to the amplitude of oscillations in the form of a Dingle-type factor,

\be
\sim \exp\left(-\frac{2 \pi m_\perp}{b \tau_{\tn{imp}} }\right),
\ee

\begin{figure*}
\begin{center}
\includegraphics[width=7.0in]{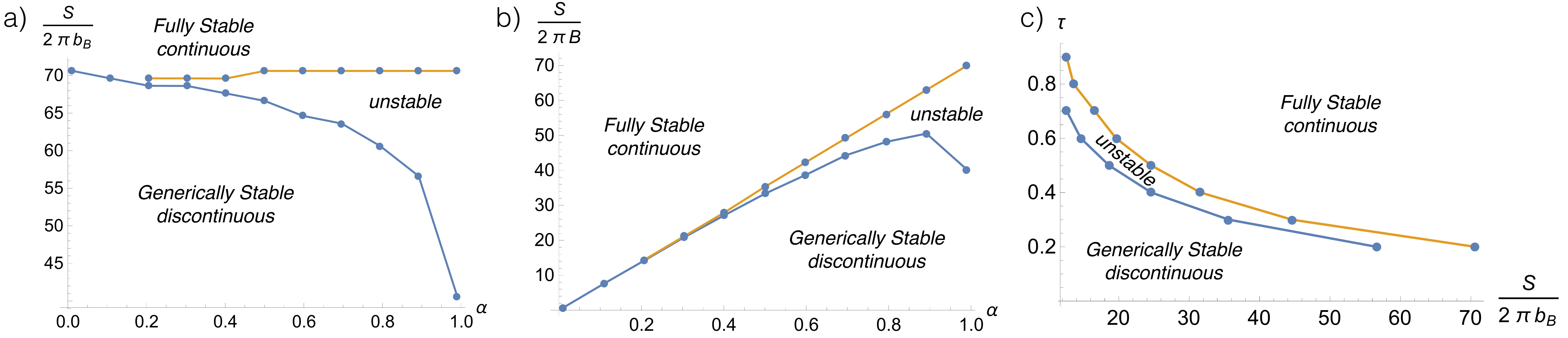}
\end{center}
\caption{(Color online) Three kinds of behavior: ``generically stable/discontinuous" regions have discontinous magnetization as function of $B$ field (see Fig.~\ref{spinon3Db}(a)) but the only thermodynamic instabilities occur at the isolated values of magnetic field associated with discontinuities;  ``fully stable continuous"  regions have continuous magnetization and no thermodynamic instabilities; ``unstable" regions have thermodynamic instabilities over finite ranges of magnetic fields (although not for every magnetic field, as stable and unstable regions alternate as the magnetic field is swept). (a) Regions at fixed temperature $\tau=0.2$ as a function of the parameter $\alpha=\chi_\varphi/(\chi_\varphi+\chi_\psi)$ controlling the effective field experienced by the neutral fermions $b_B=\alpha B$. (b) same as (a) but plotted as a function of the physical magnetic field $B$ instead of $b_B$. (c) Regions as a function of effective temperature $\tau$ and $b_B$ for fixed $\alpha\approx 0.9$. All plots are generated using the ratio $\chi_{\rm osc}/\chi_\psi=3 \sqrt{2}$ for spinless parabolic fermions and in the limit $\chi\rightarrow 0$ to allow visualization of the unstable regions. In the more realistic limit $\chi\rightarrow \infty$ the unstable region will shrink to become an increasingly thin sliver, but the generically stable/discontinuous region will remain largely unchanged.
\label{spinon3Dinst3}
}
\end{figure*}

\noindent where $\tau_{\tn{imp}}$ is the quantum lifetime of the neutral fermion as a result of scattering off impurities. Thus a simple way to estimate the impact of disorder is by absorbing this Dingle factor into the coefficient $\chi_{osc}\rightarrow \chi_{osc} \exp\left(-2 \pi m_\perp/(b_B \tau_{\tn{imp}} )\right)$ that controls the amplitude of the oscillations in our previous discussions. One interesting consequence of including this factor occurs in the two-dimensional systems discussed in Section~\ref{oscT}. Following Eq.~\eqref{DeltaBlow}, we expect that the period of oscillations at low temperatures in two-dimensions can depend on the strength of disorder and magnetic field via this factor.

There is an interesting consequence of disorder beyond the mean field level in two-dimensions. As even featureless disorder is expected to couple like a random field to the energy density, following general arguments~\cite{ImryMa,ImryWortis,AizenmanWehr}, the first order transitions that we have encountered at low temperatures in two-dimensions will be rounded into continuous transitions. However, if disorder is weak, the rounding will be mild and the transitions could in practice appear to be essentially discontinuous.

\subsection{Intermediate temperature instabilities in neutral fermi seas}\label{intermT}
 
Metals in two and three dimensions have low temperature instabilities towards phase separation. As we have seen in the previous two sections, at low temperatures fractionalized neutral Fermi seas find a way to circumvent these generic instabilities by adjusting their internal magnetic fields to find local minima of the free energy which remain thermodynamically stable. At elevated temperatures neither metals nor fractionalized Fermi seas will display instabilities simply because the oscillatory component of the magnetization will be exponentially suppressed. There remains to be explored the intermediate temperature regime at which the crossover between these two kind of behaviors happens. This will be the subject of discussion in this section. 

We will begin describing a direct numerical analysis of this intermediate regime and will later provide analytic results. We consider a simplified model with only the leading harmonic of the oscillatory part of the free energy in three dimensions. We therefore use the following total free energy of the system (including the background matter and vacuum energies):

\begin{equation}\label{f1harm}
\begin{split}
f\approx &\left(\chi+\frac{\chi_\varphi \chi_\psi}{\chi_\varphi + \chi_\psi}\right)\frac{B^2}{2}+\frac{\chi_\psi+\chi_\varphi}{2} (b-b_B)^2 \\
&- \frac{\chi_{\rm osc} b^2}{2} \left(\frac{2\pi b}{S_\perp}\right)^{1/2} \frac{\frac{\tau S_\perp}{2\pi b} }{\sinh(\frac{\tau S_\perp}{2\pi b} )} \cos\left(\frac{S_\perp}{b} \mp \frac{\pi}{4}\right).
\end{split}
\end{equation}

\noindent Here the first term is the explicit expression for $\tilde{f}_0$, which includes the energy of the vacuum and background matter in the term proportional to $\chi$. The form above is enough to capture the essential physics even at low temperatures, but more so at intermediate temperatures since the exponential temperature suppression is more pronounced for the higher harmonics. 

The task is to numerically minimize Eq.~\eqref{f1harm} as a function of $b$ for fixed $B$. Typical numerical solutions are depicted in Figures~\ref{spinon3Da} and~\ref{spinon3Db}. Figure~\ref{spinon3Da}(b) shows the typical staircase behavior of the internal magnetic field as a function of external field. As the temperature is increased this staircase smoothens and transforms into a straight line as evidenced in the low field portion of the green curve in Fig.~\ref{spinon3Da}(b). Figure~\ref{spinon3Db}(b) displays the behavior of the magnetization that results from this staircase. The magnetization is a discontinuous function of external field at low temperatures and transforms into a continuous function as temperature is increased. The low temperature discontinuities can be thought of as first order phase transitions. In particular Fig.~\ref{spinon3Db}(b) displays the second derivative of the free energy without the $\tilde{f}_0$ smooth background (see description of $\tilde{f}_0$ below Eq.~\eqref{fprime}). We notice that the second derivative has negative delta-function-like spikes at the fields for which the magnetization jumps discontinuously. Beyond a critical field (or temperature) these delta function spikes disappear and become regions with negative second derivatives of the free energy (Fig.~\ref{spinon3Db}(b)). As discussed in the introduction, a negative second derivative of the free energy indicates a thermodynamic instability towards phase separation. 

\begin{figure}[t]
	\begin{center}
		\includegraphics[width=3in]{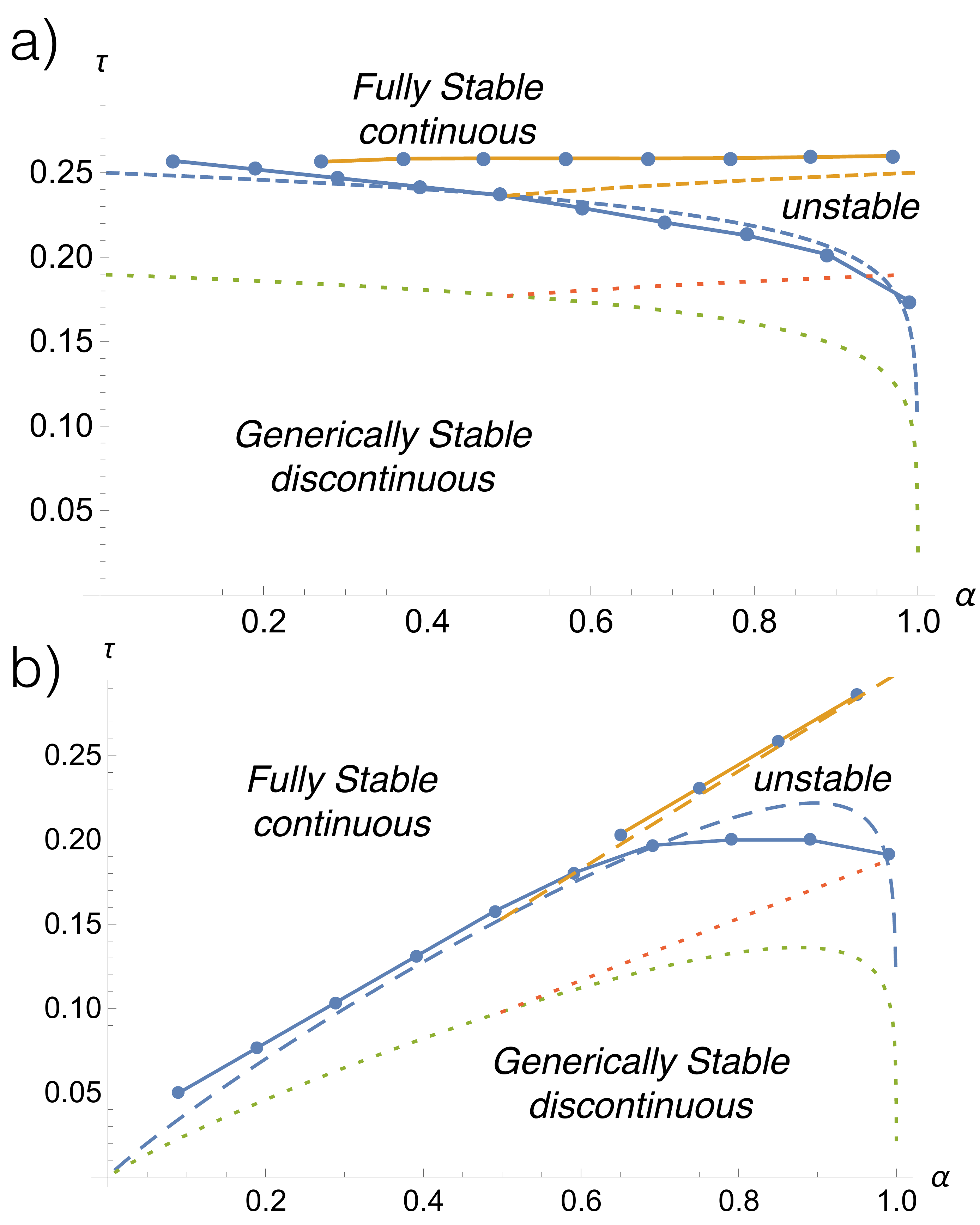}
	\end{center}
	\caption{(Color online) (a) Dimensionless temperature $\tau$ and $\alpha=\chi_\varphi/(\chi_\varphi+\chi_\psi)$ phase diagram of different regions for a constant average internal effective field $\frac{2 \pi S_\perp}{b_B}=55$. The dots joined by solid lines correspond to the regions determined from direct numerical study. The blue and orange dashed lines are the boundaries obtained from solving exactly the inequalities appearing in Eqs.~\eqref{inq1} and~\eqref{inq2} and the  green and red dotted lines are obtained from the approximations described in Eqs.~\eqref{T1approx} and~\eqref{T2app}. (b) Analogous to figure (a) but holding the physical magnetic field fixed at $\frac{2 \pi S_\perp}{B}=45$ instead of the effective field $b_B$. All plots are generated using the ratio $\chi_{\rm osc}/\chi_\psi=3 \sqrt{2}$. {}}
	\label{spinon3Dinst2}
\end{figure}

Thus we see that the fields at which the first order transitions happen are expected to turn into the regions at which thermodynamic instabilities occur over a finite region of fields, as parameters are varied (i.e field is decreased or temperature is increased). In order for a thermodynamic instability to occur, the full free energy must have a negative second derivative with respect to the external field, including the magnetic energy of vacuum. Thus, the criterion for thermodynamic stability is:

\be\label{stab}
\frac{\partial^2 f}{\partial B^2} > 0,
\ee

\noindent where we imagine that we have explicitly solved for $b$ as a function of $B$ while taking the derivatives. Notice that this second derivative will contain $\chi$, which accounts for the energy of vacuum as well as the background trivial matter. Therefore, the regions over which instabilities occur are dictated by the somewhat extrinsic parameter $\chi$.

To facilitate numerical identification, we perform the numerical analysis in the limit $\chi\rightarrow 0$, so that there is no background helping stabilize the uniform state\footnote{We are restricting to the natural case of $\chi$ being strictly positive otherwise the background ``vacuum" would have instabilities of its own.}, and will discuss later on the more realistic case of large $\chi$. Phase diagrams depicting the behavior as a function of the different dimensionless parameters of the problem are presented in Figs.~\ref{spinon3Dinst3} and~\ref{spinon3Dinst2} in the limit of $\chi\rightarrow 0$. Figure~\ref{spinon3Dinst2} shows the three kinds of regimes present in the problem. At high temperatures we have the fully stable quantum oscillations regime that resembles that of a metal but in a reduced effective field $b_B=\alpha B$. In this regime the magnetization curve is continuous and the stability criterion in Eq.\eqref{stab} is always satisfied. We label this regime as ``fully stable continuous" behavior in the plots. At low temperatures we have a regime in which the internal gauge field remains ``pinned" at certain values as the external field is swept giving rise to the staircase pattern described in Section~\ref{3DfiniteT}. In this low temperature regime instabilities occur at isolated values of the field $b_B$ for which the system transitions between different metastable states, giving rise to discontinuous jumps in the magnetization. We label this region in the plot as ``generically stable discontinuous" region. Finally there is the intermediate region between these two regimes where the stability condition in Eq.~\eqref{stab} is violated over finite ranges of magnetic field $b_B$, at which phase separation will occur. Notice that generically the system will have an alternation of regions that are stable and unstable as the magnetic field is swept in this regime. We label this region ``unstable" in the plots.

The limit $\chi\rightarrow 0$ makes the size of the unstable region as large as possible; however this exaggerates its size in comparison to the expected realistic behavior. We will now provide a quantitative description of the instability boundaries that can be used for more realistic estimates. The internal magnetic field experienced by the neutral fermions $b$ will be pinned whenever the second derivative of the smooth part with respect to $b$ is smaller than the second derivative of the oscillatory part. Thus discontinuous behavior can be estimated to occur whenever the following inequality is satisfied (assuming low fields $b\ll S_\perp$),

\be\label{inq1}
\chi_\varphi+\chi_\psi < 2 \pi^2 \chi_{osc} \left(\frac{S_\perp}{2 \pi b}\right)^{3/2} \frac{\frac{\tau S_\perp}{2\pi b} }{\sinh(\frac{\tau S_\perp}{2\pi b} )}.
\ee

\noindent The region bounded by this inequality is depicted in Fig.~\ref{spinon3Dinst2} as a dashed blue line. For the above to be satisfied, one needs that the following number be less than $\mathcal{O}(1)$:

\be\label{bO1}
\frac{\chi_\varphi+\chi_\psi}{2 \pi^2 \chi_{osc}} \left(\frac{2 \pi b_B}{S_\perp}\right)^{3/2}< \mathcal{O}(1). 
\ee

\noindent One can then replace $\frac{x}{\sinh{x}}\sim e^{-x}$ at $x\gtrsim 1$ to obtain a simple closed form expression in order to estimate the critical temperature at which discontinuities appear,

\be\label{T1approx}
T_1 \approx \frac{b_B}{2 \pi^2 m_\perp} \log \left( {\rm max}\left[1,\frac{2 \pi^2 \chi_{osc}}{\chi_\varphi+\chi_\psi} \left(\frac{S_\perp}{2 \pi b_B}\right)^{3/2}\right]\right),
\ee

\noindent which is depicted as a dotted green line in Fig.~\ref{spinon3Dinst2} that captures qualitatively well the trends and values of the numerical solution. The second temperature scale is that at which the second derivative with respect to $B$ becomes negative and generic instabilities appear whenever the following inequality is satisfied:

\begin{equation}\label{inq2}
\begin{split}
\chi+\frac{\chi_\varphi \chi_\psi}{\chi_\varphi + \chi_\psi} & <\\
& 2 \pi^2 \alpha^2 \chi_{osc} \left(\frac{S_\perp}{2 \pi b_B}\right)^{3/2} \frac{\frac{\tau S_\perp}{2\pi b_B} }{\sinh(\frac{\tau S_\perp}{2\pi b_B} )}.
\end{split}
\end{equation}

\noindent To obtain a solution to this inequality one needs that the following is satisfied:

\be
\frac{\chi+\frac{\chi_\varphi \chi_\psi}{\chi_\varphi + \chi_\psi}}{2 \pi^2 \alpha^2 \chi_{osc}} \left(\frac{2 \pi b_B}{S_\perp}\right)^{3/2}< \mathcal{O}(1). 
\ee

\noindent Using a similar approximation scheme, one obtains:

\begin{equation}\label{T2app}
\begin{split}
T_2 \approx &\frac{b_B}{2 \pi^2 m_\perp} {\rm max}\{ T_1,\\
&  \log \left( {\rm max}\left[1,\frac{2 \pi^2 \alpha^2 \chi_{osc}}{\chi+\frac{\chi_\varphi \chi_\psi}{\chi_\varphi + \chi_\psi}} \left(\frac{S_\perp}{2 \pi b_B}\right)^{3/2}\right]\right)\}.
\end{split}
\end{equation}
 
\noindent The appearance of $T_1$ in the above inequality comes from the fact that in order to use the approximation $b\approx b_B=\alpha B$ one has to assume that the critical temperature derived from the inequality~\eqref{inq2} is larger than $T_1$. Therefore, $T_2$ should be taken as the largest value between $T_1$ obtained from inequality~\eqref{inq1} and the one obtained from the inequality~\eqref{inq2}. As one approaches the metal to insulator transition $\chi_\varphi\rightarrow \infty$, $\alpha \rightarrow 1$, $b_B\rightarrow B$ and other quantities such as $\chi_\psi$ and $\chi_{osc}$ can be assumed to remain non divergent within mean field approximation. In this limit $T_1\rightarrow 0$ whereas $T_2$ becomes the value that dictates the temperature for the Condon-domain formation instability in ordinary metals:

\begin{equation}\label{T2metal}
\begin{split}
\lim_{\alpha \rightarrow 1}  & T_2 \approx \\
&  \frac{B}{2 \pi^2 m_\perp} \log \left( {\rm max}\left[1, \frac{2 \pi^2 \chi_{osc}}{\chi+\chi_\psi} \left(\frac{S_\perp}{2 \pi B}\right)^{3/2}\right]\right).\end{split}
\end{equation}

\noindent This occurs only for fields satisfying
 
\be\label{Bmetal1}
\frac{\chi+\chi_\psi}{2 \pi^2 \chi_{osc}} \left(\frac{2 \pi B}{S_\perp}\right)^{3/2} < \mathcal{O}(1). 
\ee

\noindent The maximum as a function of $B$ of Eq.~\eqref{T2metal} is achieved whenever
\be\label{Bmetal2}
B_{max} \approx \frac{S_\perp}{2\pi e} \left(\frac{2 \pi^2 \chi_{osc}}{\chi+\chi_\psi}\right)^{2/3} \sim 6 - 30\ {\rm Tesla},
\ee

\noindent where $e$ is Euler's constant, and where we used the typical values of the diamagnetic response of metals $\chi_{osc}\sim \chi_\psi \sim 10^{-6} \chi$ and typical metallic densities $n \sim 10^{28} - 10^{29} m^{-3}$. This leads to the following estimate for the maximum temperature for the phase separation instability in a typical metal

\begin{equation}\label{T2max}
\begin{split}
\lim_{\alpha \rightarrow 1}  T_2^{max} & \approx  \frac{3 \epsilon_F}{4  \pi^2 e} \left(\frac{2 \pi^2 \chi_{osc}}{\chi+\chi_\psi}\right)^{2/3} \\
& \sim 10^{-5} \epsilon_F \sim 0.4 - 3\ {\rm Kelvin}.
\end{split}
\end{equation}

\noindent for $\epsilon_F \sim 2 - 10 $ eV. These numbers are in good agreement with the parameters at which the Condon domain instabilities are typically seen~\cite{Domains99,Domains99b,gordon2003magnetic,Domains05,egorov2005condon}.

Unlike the physical magnetic field in metals, the fractionalized neutral Fermi surfaces can generically relax the global internal gauge field, reducing the region of instabilities to isolated values of the magnetic field at low temperatures, as we have described. One of the main differences between the phenomena in fractionalized states and that in metals is that the range of effective fields $b_B$ over which the discontinuous behavior occurs will become wider as we enter further into the insulating regime ($\alpha \rightarrow 0$) as indicated by Eq.~\eqref{bO1}. In both cases however the typical temperature for the appearance of instabilities is controlled by the effective cyclotron frequency $T\sim b_B/m_\perp$ of the fermions up to logarithmic corrections. Phase diagrams as a function of dimensionless temperature $\tau$ and $\alpha$ are shown in Fig.~\ref{spinon3Dinst2} in the limit $\chi\rightarrow 0$. This exaggerates the region for instabilities but allows to visualize it more easily. 

Let us discuss the more realistic limit $\chi\rightarrow \infty$ in fractionalized neutral Fermi surface states. The boundary given by $T_1$ will remain essentially unchanged because it is independent of this parameter. On the other hand the boundary given by $T_2$ will shrink and become a thin sliver surrounding the region given by $T_1$. The region bounded by $T_1$ corresponds to the behavior of the fractionalized Fermi sea that we have labeled ``generically stable discontinuous" in the plots. The region of effective fields $b_B$ for this regime is expected to be considerably larger than that of the conventional Condon Domain formation in ordinary three dimensional metals. The critical temperature for a given field is still controlled by the effective cyclotron frequency of the neutral fermions $T_1 \sim b_B/m_\perp$, up to logarithmic corrections. Unlike metals, the region of instability in magnetic fields can even extend into the deep quantized regime. To appreciate this more directly we can recast the inequality~\eqref{bO1} that describes the field region of ``generically stable discontinuous" behavior as follows

\be\label{BT1}
B\lesssim \frac{S_\perp}{2 \pi} \frac{(1-\alpha)^{2/3}}{\alpha} \left(\frac{2 \pi^2 \chi_{osc}}{\chi_{\psi}}\right)^{2/3}.
\ee

\noindent Since the ratio $\chi_{osc}/\chi_{\psi}$ is expected to be of order $1$, this extends well into the deep quantized regime of $B\sim S_\perp/\alpha$ as we move deeper into the insulating limit, $\alpha\rightarrow 0$ (contrast this range of fields with that in metals in Eq.~\eqref{Bmetal1} and~\eqref{Bmetal2}). In this ultra-quantum regime the critical temperatures for the appearance of generically stable but discontinuous behavior are on the order of the Fermi temperature. In fact the maximum temperature,  $T_1$, is achieved at a field on the order of that in Eq.~\eqref{BT1} and has typical values of the form

\begin{equation}\label{T1max}
\begin{split}
T_1^{max} & \approx  \frac{3 \epsilon_F}{4 \pi^2 e} \left(\frac{2 \pi^2 \chi_{osc}}{\chi_\varphi+\chi_\psi}\right)^{2/3} \\
& \sim 0.5 \epsilon_F,
\end{split}
\end{equation}

\noindent where in the second line we did the estimate considering the limit $\chi_\varphi\rightarrow 0$. We caution however that to achieve the ultra-quantum regime of the neutral fermions becomes harder as one goes deeper into the insulator, since the effective field experienced by the fermions is reduced by the factor $\alpha$. Moreover, our theory is developed under the assumption of small fields, $b_B\ll S_\perp$, so strictly speaking we are naively extrapolating outside the regime of control of the theory, but the qualitative point that we wish to emphasize is the trend that the critical temperatures separating the fully stable uniform phases from the phase separation instability regimes will become typically much higher in fractionalized Fermi seas as we move deeper into the insulator.

\begin{table*}
\caption{Summary of the behavior of the period $\Delta (1/B)$ (in units of $2\pi/S_\perp$) and amplitude $4\pi \delta M_{osc}$ (in units of $\chi_\varphi S_\perp/(2\pi)$) of the magnetization oscillations in two-dimensional and three-dimensional fractionalized fermi seas. Here $\alpha=\chi_\varphi/(\chi_\psi+\chi_\varphi)$ and $\omega_\psi=\alpha B/m_\perp$ (units $e=c=\hbar=1$). $\chi_{osc}$ controls the amplitude of the oscillations; in the clean limit  of spinless Galilean fermions, it is given by $\chi_{osc}=12 \chi_\psi/\pi^2$ in two-dimensions and $\chi_{osc}=3 \sqrt{2} \chi_\psi$ in three-dimensions. The impact of a short-range disorder potential can be captured by adding a Dingle factor into this coefficient $\chi_{osc}\rightarrow \chi_{osc} \exp\left(-2 \pi m_\perp/(\alpha B \tau_{\tn{imp}} )\right)$.}  
\begin{tabular}{c  c  c} 
\hline\hline 
 & 2D & 3D \\ [0.5ex] 
\hline 
period $T\ll \omega_\psi$ & $\frac{\chi_\varphi}{\chi_\psi+\chi_\varphi-\pi^2 \chi_{osc}/12}$ & $\alpha$ \\ [0.5ex]
period $T\gtrsim \omega_\psi$  & $\alpha$ &  $\alpha$ \\ [1ex] 
Amplitude $T\ll \omega_\psi$ & $\left(\frac{\chi _{\varphi }}{\chi _{\varphi }+\chi _{\psi }-\pi^2 \chi_{osc}/12} \frac{2\pi B}{S}\right)^2$ & $\left(\frac{2\pi \alpha B}{S_\perp}\right)^2$ \\ [0.5ex]
Amplitude $T\gtrsim \omega_\psi$  & $\frac{2 \pi \alpha \chi_{\rm osc}}{\chi_\varphi}  \frac{\frac{2 \pi^2 T}{\omega_\psi}}{ \sinh \left(\frac{2 \pi^2 T}{\omega_\psi}\right)}$ &  $\frac{2 \pi \alpha \chi_{\rm osc}}{\chi_\varphi} \left(\frac{2\pi \alpha B}{S_\perp}\right)^{1/2} \frac{\frac{2 \pi^2 T}{\omega_\psi}}{ \sinh \left(\frac{2 \pi^2 T}{\omega_\psi}\right)}$ \\  [1ex] 
\hline\hline 
\end{tabular}
\label{sumtab} 
\end{table*}

We close this section by noting that Condon domain formation in metals occurs at temperatures of the order of Kelvins which are well above those for which the de-Haas van-Alphen effect is still visible. Thus in practice the domain formation does not affect the observation of the de-Haas van-Alphen effect in metals, but changes the harmonic content of the shape of the measured magnetization as a function of field. In the semiclassical regime, we expect that the same will largely hold in fractionalized Fermi seas, namely the domain formation will not prevent the observation of magnetization oscillations in systems with neutral Fermi surfaces, but will simply change the precise shape of the magnetization curve, especially its higher harmonics, just like in metals. Perhaps the fact that this phenomenon is expected to survive to higher temperatures and magnetic fields might facilitate the observation of these domains in fractionalized Fermi seas with a sizable value of the parameter $\alpha$.

\section{Resistivity oscillations of fractionalized Fermi sea}\label{resistivity}

In this section, we will consider the possibility that electrical insulators with emergent neutral Fermi surfaces display quantum oscillations in their resistivity at finite temperature. As we will see, these oscillations will be superimposed on the activated temperature dependence, characteristic of the insulating behavior of such phases. We will discuss suitable conditions that will make more likely their observation in actual materials.

Even though the phase under consideration is an insulator, it will generically have a finite resistivity at finite temperature. The Ioffe-Larkin rule~\cite{IL} states that the resistivities, rather than the conductivities, of the Fermionic and bosonic partons add up to produce the physical net resistivity of the system,

\be
\rho=\rho_\psi+\rho_\varphi.
\ee

\noindent Since the boson is gapped one expects an Arrhenius-type activated behavior of its resistivity with an energy scale controlled by the charge gap ($\Delta$)

\be
\rho_\varphi \approx \rho_\varphi^0 e^{\frac{\Delta}{T}}.
\ee

\noindent The fermion, however, experiences an effective magnetic field and hence will develop an oscillatory component in the resistivity. Detailed theories of quantum oscillation of resistivities are rather involved. Here we will follow a simple approach that captures correctly the order of magnitude of the effect in metals. In metals the main effect of the magnetic field is to change the scattering rate off impurities. Such rate is proportional to the density of states and hence oscillates as a result of the Landau quantization~\cite{abrikosov}. Following the analogous derivation for a metal detailed in Ref.~\cite{abrikosov}, one obtains that the resistivity will have an oscillatory component given by,

\begin{equation}\label{rhopsi}
\begin{split}
&\frac{\delta \rho_\psi^{\rm osc}}{\rho_\psi^0} \sim \frac{1}{\nu_F}\left(\frac{m_\perp b }{S_\perp}\right)^2\frac{\partial^2 f_{osc}}{\partial b^2}\approx\\
& \frac{\chi_{\rm osc}}{24 \chi_\psi} \left(\frac{2\pi b}{S_\perp}\right)^{1/2} \sum_{k=1}^\infty \frac{(-1)^{k+1}}{k^{1/2}} \frac{\frac{k S \tau }{2 \pi  b}}{ \sinh \left(\frac{k S \tau }{2 \pi  b}\right)} \cos \left(\frac{k S}{b} \mp \frac{\pi}{4}\right),
\end{split}
\end{equation}

\noindent where $\rho_\psi^0$ is the resistivity of the fermions at zero effective field, $\delta \rho_\psi^{\rm osc}$ is the oscillatory component of the resistivity, and $f_{\rm osc}$ is given in Eq.~\eqref{fosc3d}. As we have seen in the previous section, the internal magnetic field experienced by the Fermions is $b_B=\alpha B$, in the high temperature regime $T\gtrsim \omega_\psi$. However, apart from the rescaling of the effective field, the functional from of the oscillatory component of the resistivity is essentially the same as that in metals. At lower temperatures the internal magnetic field will follow a stair-case type behavior around the average value $b_B=\alpha B$, due to the pinning of the internal field at the local minima of the free energy described in Section.~\ref{3DfiniteT}. This will typically lead to a form of the resistivity oscillations that is more anharmonic than that of metals at low temperatures. In spite of this, at the mean field level, the period and the amplitude of the oscillations is expected to have the same functional form as that in metals, even at low temperatures ($T\ll \omega_\psi$). Therefore the amplitude of the oscillatory component of the resistivity is expected to follow an expression analogous to the Lifshitz-Kosevich form,

\be
\frac{\delta \rho_\psi^{\rm osc}}{\rho} \sim \frac{\rho_\psi^0 }{\rho_\varphi^0 e^{\frac{\Delta}{T}}}\left(\frac{2\pi b_B}{S_\perp}\right)^{1/2} \frac{\frac{S_\perp \tau }{2 \pi  b_B}}{ \sinh \left(\frac{S_\perp \tau }{2 \pi  b_B}\right)} e^{-\frac{2 \pi m_\perp}{b_B \tau_{\tn{imp}} }}.
\ee

\noindent Here we have added a Dingle factor to capture the impact of disorder on the amplitude of resistivity oscillations. Notice that the oscillatory component is exponentially small relative to the total resistance due to the temperature activated resistive background. In this sense, we expect that the effect will be hard to detect in strongly insulating materials with a large charge gap. In addition the Dingle factor highlights that disorder will contribute further to reduce the amplitude of the oscillatory component, eventually suppressing the effect at strong disorder. Therefore the ideal materials for the observation of this effect will lie in a ``sweet spot" that corresponds to those which have a relatively small charge gap and thus are not strongly insulating, while being sufficiently clean for the amplitude of the oscillations to be visible.

However, we wish to emphasize that the above is the dependence based on the mean field theory under the assumption that the effective magnetic field only affects the fermion scattering by changing their density of states. As we have discussed in Section~\ref{intermT} there is a large region in which the solutions with uniform effective field $b$ are unstable to states with inhomogeneous values of $b$. We believe that, in the semiclassical regime $b\ll S_\perp$, this will not have a substantial effect on the major features of the oscillations because the fluctuations of the effective field are between the values associated with consecutive minima of the free energy, and therefore should not change dramatically the disorder landscape. Thus we expect that this effect will not hinder the observation of the resistivity oscillations but will change details such as the precise harmonic content of the oscillations and perhaps even enhance the scattering off domain walls leading to larger amplitude of the resistivity oscillations. This expectation is partly based on the rule of thumb that such domain formation in metals does not affect the essential observation of the Shubnikov-de Haas effect.

We would like to mention a caveat that applies to the mixed valence insulators for which we have conjectured the presence of the composite exciton Fermi liquid~\cite{CSS}. These materials are known to possess a metallic surface~\cite{Li14,SS15} which has even been argued to have a topological origin~\cite{Coleman1}. The net resistance of these materials can be modeled as arising from two resistors in parallel accounting for the surface and the bulk conduction,

\be\label{series}
R_{\rm tot}=\frac{R_{\rm surf} R_{\rm bulk}}{R_{\rm surf} + R_{\rm bulk}}.
\ee

\noindent At temperatures well below the crossover value at which the resistivity ceases to have the activated Arrhenius-type behavior and saturates into a metallic surface dominated regime, most of the electrical current will flow through the surface because in this regime $R_{\rm bulk}\gg R_{\rm surf}$. The bulk resistance can then be split as the sum of the resistance of the boson and the fermions,  following the Ioffe-Larkin rule, as $R_{\rm bulk}=R_\varphi+R_\psi$. Taking the limit $R_\varphi\rightarrow \infty$ in Eq.~\eqref{series} leads to the following leading behavior of the net resistance,

\be\label{series}
R_{\rm tot}\approx R_{\rm surf}- e^{-\frac{\Delta}{T}} \frac{R_{\rm surf}^2}{R^0_{\varphi}}+e^{-2 \frac{\Delta}{T}}  \frac{R_{\psi} R_{\rm surf}^2+R_{\rm surf}^3}{(R^0_{\varphi})^2},
\ee

\noindent where we have used an Arrhenius form for the boson resistance $R_\varphi=R^0_{\varphi} e^{\frac{\Delta}{T}}$ to highlight its temperature dependence. Therefore we see that the contribution from the bulk quantum oscillations of the neutral fermions is effectively exponentially suppressed when the temperature is well below the value for which the resistance saturates, simply because the electrical current is shunted through surface of the material. Therefore, the effect we have mind should be extracted from temperatures higher than this crossover temperature, which in principle, can be pushed to lower temperatures by increasing the size of the sample as this enhances its volume to surface ratio.

To summarize this section, we have seen that the fractionalized Fermi sea will behave as an insulator in the sense that the resistivity will increase as the temperature is lowered, but, unlike conventional insulators, this smooth rise will be accompanied by an oscillatory component that resembles that of a metal. The physical origin of this effect can be traced back to the fact that the finite amount of thermally induced bosonic carriers will self-consistently set an internal electric field when a steady current flows, which will itself act on the fermions. This effect might be most visible in systems where the charge gap of the insulator is not so large so as to completely overwhelm the oscillatory component arising from the neutral Fermi sea.

\section{summary and discussion}\label{sum}

We have developed a quantitative theory of quantum oscillations in the magnetization of fractionalized neutral Fermi seas in response to external magnetic fields, with special focus on the temperature dependence of these oscillations. Table~\ref{sumtab} summarizes some of our main formulae describing magnetization oscillations in two- and three-dimensional fractionalized neutral Fermi seas. In these systems, the neutral fermions experience an internal magnetic field of the emergent gauge field that is set in self-consistently due to the diamagnetism of the gapped charge carrying degrees of freedom. 

Quite generally, the oscillations display distinct behavior depending on whether the temperature is lower or higher than the effective cyclotron energy of the neutral fermions. At high temperatures the oscillations resemble those of a metal, albeit in an effective field that has a different strength from the physical magnetic field inside the sample. At low temperatures the system displays a multiplicity of metastable states and quantum oscillations can be viewed as a series of phase transitions between these states. This phenomenon resembles Condon domain formation in ordinary metals, but it is distinct from it because of the ability of the internal gauge field to adjust itself to minimize the free energy. As a result, the instabilities at low temperatures occur only at isolated values of the external magnetic field, rather than over finite ranges as is the case in ordinary metals. We believe that even if experiments reach such ultra-low temperature regime this will not hinder the essential observation of quantum oscillations but will affect details, such as the precise amplitude of different harmonics, as is the case in metals. 

We have also discussed an analogue of the Shubnikov-de Haas oscillations of the resistivity in metals. In the present case this is possible because of the thermal activation of current carrying quasiparticles. Following the Ioffe-Larkin rule that dictates the addition of the resistivities of the partons, we conclude that the fractionalized neutral Fermi sea will display an oscillatory resistivity superimposed with the more conventional activated resistivity behavior that characterizes insulators.

\begin{figure}
\begin{center}
\includegraphics[width=0.9\columnwidth]{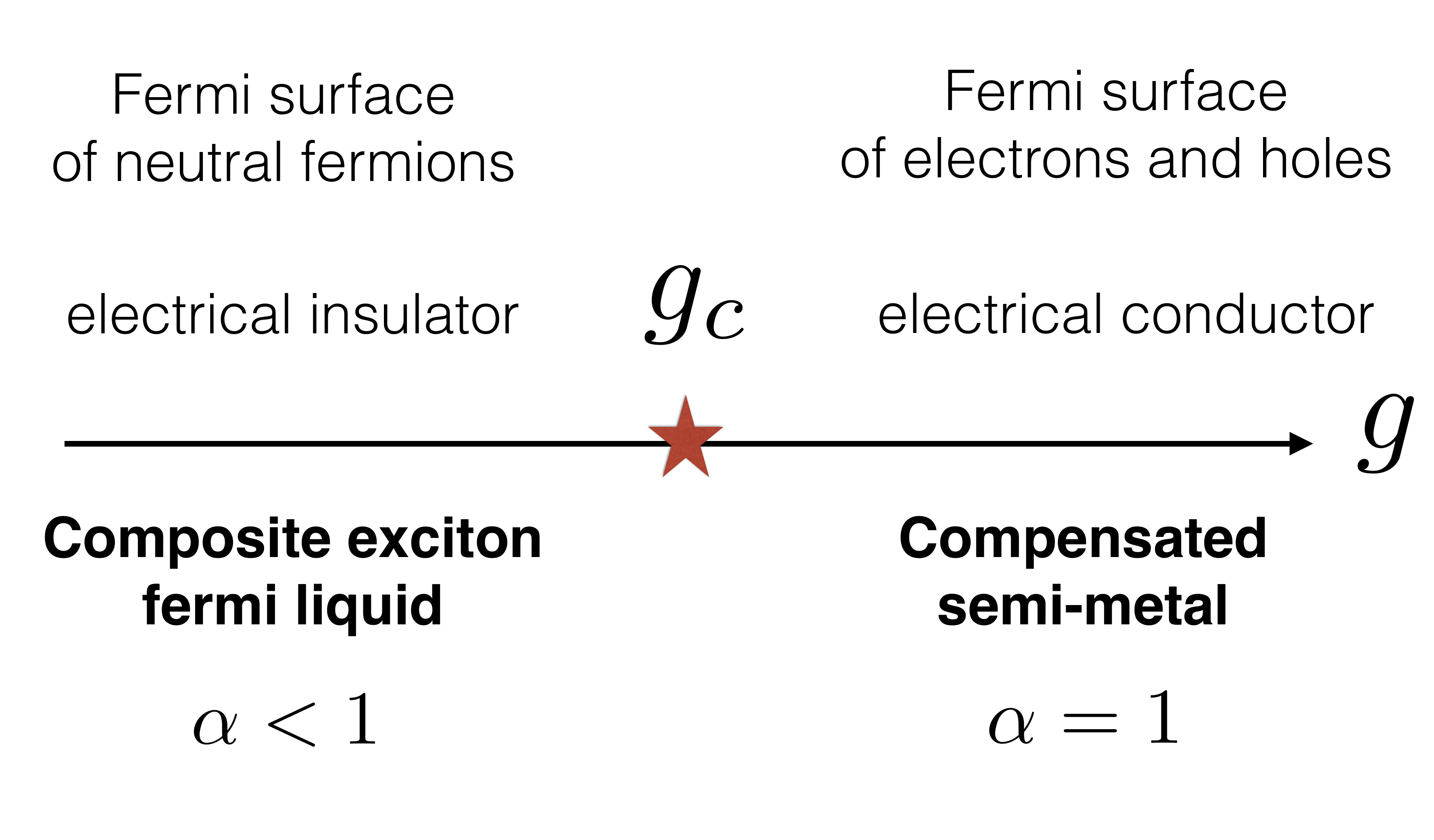}
\end{center}
\caption{(Color online) The composite exciton fermi liquid state can have a direct metal to insulator transtion to an ordinary compensated semimetal. The proximity of a material with the composite exciton fermi liquid phase to such a critical point will enhance the possibility to observe quantum oscillations by enhancing the effective field of the fermions and reducing the activated resistive background of the insulator. An analogous statement can be made for a system with a spinon fermi surface state which is in proximity to a metal to insulator transition quantum critical point.}
\label{critical}
\end{figure}

\subsection{Connections to materials}\label{exptsum}

As we have seen, the composite exciton Fermi liquid phase that we have proposed to arise in mixed valence insulators~\cite{CSS} supports quantum oscillations as a result of the fact that an external magnetic field induces a finite value for the emergent magnetic field to which the neutral fermions couple. The parameter $\alpha$ that controls the strength of the internal magnetic field experienced by the fermions is expected to approach $1$ as the system approaches a critical point between the insulator and a compensated semi-metal (see Fig.~\ref{critical}). Certain mixed valence insulators might lie closer to such a critical point, enhancing the field experienced by the neutral fermion and thus enhancing the possibility to observe quantum oscillations. In fact, more recent measurements on a mixed valence insulator compound different from SmB$_6$, display clear bulk quantum oscillations and have a finite intercept for the ratio of $\kappa_{xx}/T$~\cite{MatsudaLi} down to the lowest measurable temperatures, in a clear indication of the formation of a Fermi surface of neutral fermions. As we have discussed, the observation of resistivity oscillations would require to meet certain suitable conditions--- the material should not be strongly insulating or else it will be hard to detect the oscillations on top of a large activated resistive background and at the same time it should be sufficiently clean as for the oscillations to remain sizable. Therefore the proximity to a critical point separating the composite exciton fermi liquid from a compensated semimetal (see Fig.~\ref{critical}) will also enhance the chance to observe resistance oscillations because of the reduction of the insulating charge gap. One important caveat for mixed valence insulators with metallic surfaces is that the resistivity measurements should be performed above the  temperature where the resistivity saturates due to surface dominated transport. Observation of resistance oscillations that meet these conditions would provide independent evidence of the presence of bulk oscillations in these materials.

Resistivity oscillations would also be an important tool in the elucidation of the nature of the quantum spin liquid phases in the organic materials. Importantly, in these materials the metal to insulator transition can be driven in a clean fashion by applying pressure~\cite{pressure1996,pressure2005,pressure2015,Kanoda2017}. A natural question that arises is: what is the fate of Shubnikov-de Haas effect that is present in the metal as one tunes the pressure to drive the metal to insulator transition? Given that the metal-insulator transition is either continuous or very weakly first order~\cite{Kanoda2017}, we expect that the Shubnikov-de Haas oscillations will persist into the insulating phase. Previous experimental attempts to detect quantum oscillations in these materials were at ambient pressure and studied the magnetization. However no oscillations were detected~\cite{watanabe2012novel}. This suggests that the parameter $\alpha$ may be too small at ambient pressure. Based on the considerations in this paper we advocate searching for quantum oscillations in the resistivity under pressure close to the metal-insulator transition where $\alpha$ will approach $1$. We expect that even on the insulating side there will be quantum oscillations superimposed on the activated resistivity as described in this paper. The pressure induced tunability of the charge gap and the high quality of the organic materials might facilitate finding the ``sweet spot" to observe the resistivity oscillations that we have described.

\acknowledgements

We thank Piers Coleman, Lu Li, Yuji Matsuda, Olexei Motrunich, Suchitra Sebastian, and Takasada Shibauchi for many stimulating discussions. D.C. is supported by a postdoctoral fellowship from the Gordon and Betty Moore Foundation, under the EPiQS initiative, Grant GBMF-4303, at MIT. I.S. is supported by the MIT Pappalardo Fellowship. T.S. is supported by a US Department of Energy grant DE-SC0008739, and in part by a Simons Investigator award from the Simons Foundation.


\bibliographystyle{apsrev4-1_custom}
%


%
%
%
%
%
%


\begin{thebibliography}{48}%
\makeatletter
\providecommand \@ifxundefined [1]{%
 \@ifx{#1\undefined}
}%
\providecommand \@ifnum [1]{%
 \ifnum #1\expandafter \@firstoftwo
 \else \expandafter \@secondoftwo
 \fi
}%
\providecommand \@ifx [1]{%
 \ifx #1\expandafter \@firstoftwo
 \else \expandafter \@secondoftwo
 \fi
}%
\providecommand \natexlab [1]{#1}%
\providecommand \enquote  [1]{``#1''}%
\providecommand \bibnamefont  [1]{#1}%
\providecommand \bibfnamefont [1]{#1}%
\providecommand \citenamefont [1]{#1}%
\providecommand \href@noop [0]{\@secondoftwo}%
\providecommand \href [0]{\begingroup \@sanitize@url \@href}%
\providecommand \@href[1]{\@@startlink{#1}\@@href}%
\providecommand \@@href[1]{\endgroup#1\@@endlink}%
\providecommand \@sanitize@url [0]{\catcode `\\12\catcode `\$12\catcode
  `\&12\catcode `\#12\catcode `\^12\catcode `\_12\catcode `\%12\relax}%
\providecommand \@@startlink[1]{}%
\providecommand \@@endlink[0]{}%
\providecommand \url  [0]{\begingroup\@sanitize@url \@url }%
\providecommand \@url [1]{\endgroup\@href {#1}{\urlprefix }}%
\providecommand \urlprefix  [0]{URL }%
\providecommand \Eprint [0]{\href }%
\providecommand \doibase [0]{http://dx.doi.org/}%
\providecommand \selectlanguage [0]{\@gobble}%
\providecommand \bibinfo  [0]{\@secondoftwo}%
\providecommand \bibfield  [0]{\@secondoftwo}%
\providecommand \translation [1]{[#1]}%
\providecommand \BibitemOpen [0]{}%
\providecommand \bibitemStop [0]{}%
\providecommand \bibitemNoStop [0]{.\EOS\space}%
\providecommand \EOS [0]{\spacefactor3000\relax}%
\providecommand \BibitemShut  [1]{\csname bibitem#1\endcsname}%
\let\auto@bib@innerbib\@empty
\bibitem [{\citenamefont {Motrunich}(2006)}]{Motrunich}%
  \BibitemOpen
  \bibfield  {author} {\bibinfo {author} {\bibfnamefont {O.~I.}\ \bibnamefont
  {Motrunich}},\ }\bibfield  {title} {\enquote {\bibinfo {title} {Orbital
  magnetic field effects in spin liquid with spinon fermi sea: Possible
  application to
  $\ensuremath{\kappa}\text{\ensuremath{-}}{(\mathrm{ET})}_{2}{\mathrm{cu}}_{2}{(\mathrm{C}\mathrm{N})}_{3}$},}\
  }\href {\doibase 10.1103/PhysRevB.73.155115} {\bibfield  {journal} {\bibinfo
  {journal} {Phys. Rev. B}\ }\textbf {\bibinfo {volume} {73}},\ \bibinfo
  {pages} {155115} (\bibinfo {year} {2006})}\BibitemShut {NoStop}%
\bibitem [{\citenamefont {Shimizu}\ \emph {et~al.}(2003)\citenamefont
  {Shimizu}, \citenamefont {Miyagawa}, \citenamefont {Kanoda}, \citenamefont
  {Maesato},\ and\ \citenamefont {Saito}}]{Shimizu03}%
  \BibitemOpen
  \bibfield  {author} {\bibinfo {author} {\bibfnamefont {Y.}~\bibnamefont
  {Shimizu}}, \bibinfo {author} {\bibfnamefont {K.}~\bibnamefont {Miyagawa}},
  \bibinfo {author} {\bibfnamefont {K.}~\bibnamefont {Kanoda}}, \bibinfo
  {author} {\bibfnamefont {M.}~\bibnamefont {Maesato}}, \ and\ \bibinfo
  {author} {\bibfnamefont {G.}~\bibnamefont {Saito}},\ }\bibfield  {title}
  {\enquote {\bibinfo {title} {Spin liquid state in an organic mott insulator
  with a triangular lattice},}\ }\href {\doibase 10.1103/PhysRevLett.91.107001}
  {\bibfield  {journal} {\bibinfo  {journal} {Phys. Rev. Lett.}\ }\textbf
  {\bibinfo {volume} {91}},\ \bibinfo {pages} {107001} (\bibinfo {year}
  {2003})}\BibitemShut {NoStop}%
\bibitem [{\citenamefont {Yamashita}\ \emph {et~al.}(2008)\citenamefont
  {Yamashita}, \citenamefont {Nakazawa}, \citenamefont {Oguni}, \citenamefont
  {Oshima}, \citenamefont {Nojiri}, \citenamefont {Shimizu}, \citenamefont
  {Miyagawa},\ and\ \citenamefont {Kanoda}}]{Yamashita08}%
  \BibitemOpen
  \bibfield  {author} {\bibinfo {author} {\bibfnamefont {S.}~\bibnamefont
  {Yamashita}}, \bibinfo {author} {\bibfnamefont {Y.}~\bibnamefont {Nakazawa}},
  \bibinfo {author} {\bibfnamefont {M.}~\bibnamefont {Oguni}}, \bibinfo
  {author} {\bibfnamefont {Y.}~\bibnamefont {Oshima}}, \bibinfo {author}
  {\bibfnamefont {H.}~\bibnamefont {Nojiri}}, \bibinfo {author} {\bibfnamefont
  {Y.}~\bibnamefont {Shimizu}}, \bibinfo {author} {\bibfnamefont
  {K.}~\bibnamefont {Miyagawa}}, \ and\ \bibinfo {author} {\bibfnamefont
  {K.}~\bibnamefont {Kanoda}},\ }\bibfield  {title} {\enquote {\bibinfo {title}
  {Thermodynamic properties of a spin-1/2 spin-liquid state in a
  {$[$}kappa{$]$}-type organic salt},}\ }\href
  {http://dx.doi.org/10.1038/nphys942} {\bibfield  {journal} {\bibinfo
  {journal} {Nat Phys}\ }\textbf {\bibinfo {volume} {4}},\ \bibinfo {pages}
  {459} (\bibinfo {year} {2008})}\BibitemShut {NoStop}%
\bibitem [{\citenamefont {Yamashita}\ \emph {et~al.}(2010)\citenamefont
  {Yamashita}, \citenamefont {Nakata}, \citenamefont {Senshu}, \citenamefont
  {Nagata}, \citenamefont {Yamamoto}, \citenamefont {Kato}, \citenamefont
  {Shibauchi},\ and\ \citenamefont {Matsuda}}]{Yamashita10}%
  \BibitemOpen
  \bibfield  {author} {\bibinfo {author} {\bibfnamefont {M.}~\bibnamefont
  {Yamashita}}, \bibinfo {author} {\bibfnamefont {N.}~\bibnamefont {Nakata}},
  \bibinfo {author} {\bibfnamefont {Y.}~\bibnamefont {Senshu}}, \bibinfo
  {author} {\bibfnamefont {M.}~\bibnamefont {Nagata}}, \bibinfo {author}
  {\bibfnamefont {H.~M.}\ \bibnamefont {Yamamoto}}, \bibinfo {author}
  {\bibfnamefont {R.}~\bibnamefont {Kato}}, \bibinfo {author} {\bibfnamefont
  {T.}~\bibnamefont {Shibauchi}}, \ and\ \bibinfo {author} {\bibfnamefont
  {Y.}~\bibnamefont {Matsuda}},\ }\bibfield  {title} {\enquote {\bibinfo
  {title} {Highly mobile gapless excitations in a two-dimensional candidate
  quantum spin liquid},}\ }\href {\doibase 10.1126/science.1188200} {\bibfield
  {journal} {\bibinfo  {journal} {Science}\ }\textbf {\bibinfo {volume}
  {328}},\ \bibinfo {pages} {1246} (\bibinfo {year} {2010})},\ \Eprint
  {http://arxiv.org/abs/http://science.sciencemag.org/content/328/5983/1246.full.pdf}
  {http://science.sciencemag.org/content/328/5983/1246.full.pdf} \BibitemShut
  {NoStop}%
\bibitem [{\citenamefont {Yamashita}\ \emph {et~al.}(2011)\citenamefont
  {Yamashita}, \citenamefont {Yamamoto}, \citenamefont {Nakazawa},
  \citenamefont {Tamura},\ and\ \citenamefont {Kato}}]{Yamashita11}%
  \BibitemOpen
  \bibfield  {author} {\bibinfo {author} {\bibfnamefont {S.}~\bibnamefont
  {Yamashita}}, \bibinfo {author} {\bibfnamefont {T.}~\bibnamefont {Yamamoto}},
  \bibinfo {author} {\bibfnamefont {Y.}~\bibnamefont {Nakazawa}}, \bibinfo
  {author} {\bibfnamefont {M.}~\bibnamefont {Tamura}}, \ and\ \bibinfo {author}
  {\bibfnamefont {R.}~\bibnamefont {Kato}},\ }\bibfield  {title} {\enquote
  {\bibinfo {title} {Gapless spin liquid of an organic triangular compound
  evidenced by thermodynamic measurements},}\ }\href
  {http://dx.doi.org/10.1038/ncomms1274} {\bibfield  {journal} {\bibinfo
  {journal} {Nature Communications}\ }\textbf {\bibinfo {volume} {2}},\
  \bibinfo {pages} {275 EP } (\bibinfo {year} {2011})}\BibitemShut {NoStop}%
\bibitem [{\citenamefont {Lee}\ and\ \citenamefont
  {Lee}(2005{\natexlab{a}})}]{PAL05}%
  \BibitemOpen
  \bibfield  {author} {\bibinfo {author} {\bibfnamefont {S.-S.}\ \bibnamefont
  {Lee}}\ and\ \bibinfo {author} {\bibfnamefont {P.~A.}\ \bibnamefont {Lee}},\
  }\bibfield  {title} {\enquote {\bibinfo {title} {U(1) gauge theory of the
  hubbard model: Spin liquid states and possible application to
  $\ensuremath{\kappa}\mathrm{\text{\ensuremath{-}}}(\mathrm{BEDT}\mathrm{\text{\ensuremath{-}}}\mathrm{TTF}{)}_{2}{\mathrm{cu}}_{2}(\mathrm{CN}{)}_{3}$},}\
  }\href {\doibase 10.1103/PhysRevLett.95.036403} {\bibfield  {journal}
  {\bibinfo  {journal} {Phys. Rev. Lett.}\ }\textbf {\bibinfo {volume} {95}},\
  \bibinfo {pages} {036403} (\bibinfo {year} {2005}{\natexlab{a}})}\BibitemShut
  {NoStop}%
\bibitem [{\citenamefont {Motrunich}(2005{\natexlab{a}})}]{OM05}%
  \BibitemOpen
  \bibfield  {author} {\bibinfo {author} {\bibfnamefont {O.~I.}\ \bibnamefont
  {Motrunich}},\ }\bibfield  {title} {\enquote {\bibinfo {title} {Variational
  study of triangular lattice spin-$1/2$ model with ring exchanges and spin
  liquid state in
  $\ensuremath{\kappa}\text{\ensuremath{-}}{(\mathrm{ET})}_{2}{\mathrm{cu}}_{2}{(\mathrm{CN})}_{3}$},}\
  }\href {\doibase 10.1103/PhysRevB.72.045105} {\bibfield  {journal} {\bibinfo
  {journal} {Phys. Rev. B}\ }\textbf {\bibinfo {volume} {72}},\ \bibinfo
  {pages} {045105} (\bibinfo {year} {2005}{\natexlab{a}})}\BibitemShut
  {NoStop}%
\bibitem [{\citenamefont {Lee}\ and\ \citenamefont
  {Lee}(2005{\natexlab{b}})}]{Lee&Lee}%
  \BibitemOpen
  \bibfield  {author} {\bibinfo {author} {\bibfnamefont {S.-S.}\ \bibnamefont
  {Lee}}\ and\ \bibinfo {author} {\bibfnamefont {P.~A.}\ \bibnamefont {Lee}},\
  }\bibfield  {title} {\enquote {\bibinfo {title} {U(1) gauge theory of the
  hubbard model: Spin liquid states and possible application to
  $\ensuremath{\kappa}\mathrm{\text{\ensuremath{-}}}(\mathrm{BEDT}\mathrm{\text{\ensuremath{-}}}\mathrm{TTF}{)}_{2}{\mathrm{cu}}_{2}(\mathrm{CN}{)}_{3}$},}\
  }\href {\doibase 10.1103/PhysRevLett.95.036403} {\bibfield  {journal}
  {\bibinfo  {journal} {Phys. Rev. Lett.}\ }\textbf {\bibinfo {volume} {95}},\
  \bibinfo {pages} {036403} (\bibinfo {year} {2005}{\natexlab{b}})}\BibitemShut
  {NoStop}%
\bibitem [{\citenamefont
  {Motrunich}(2005{\natexlab{b}})}]{MotrunichVariational}%
  \BibitemOpen
  \bibfield  {author} {\bibinfo {author} {\bibfnamefont {O.~I.}\ \bibnamefont
  {Motrunich}},\ }\bibfield  {title} {\enquote {\bibinfo {title} {Variational
  study of triangular lattice spin-$1?2$ model with ring exchanges and spin
  liquid state in
  $\ensuremath{\kappa}\text{\ensuremath{-}}{(\mathrm{ET})}_{2}{\mathrm{cu}}_{2}{(\mathrm{CN})}_{3}$},}\
  }\href {\doibase 10.1103/PhysRevB.72.045105} {\bibfield  {journal} {\bibinfo
  {journal} {Phys. Rev. B}\ }\textbf {\bibinfo {volume} {72}},\ \bibinfo
  {pages} {045105} (\bibinfo {year} {2005}{\natexlab{b}})}\BibitemShut
  {NoStop}%
\bibitem [{\citenamefont {Li}\ \emph {et~al.}(2014)\citenamefont {Li},
  \citenamefont {Xiang}, \citenamefont {Yu}, \citenamefont {Asaba},
  \citenamefont {Lawson}, \citenamefont {Cai}, \citenamefont {Tinsman},
  \citenamefont {Berkley}, \citenamefont {Wolgast}, \citenamefont {Eo},
  \citenamefont {Kim}, \citenamefont {Kurdak}, \citenamefont {Allen},
  \citenamefont {Sun}, \citenamefont {Chen}, \citenamefont {Wang},
  \citenamefont {Fisk},\ and\ \citenamefont {Li}}]{Li14}%
  \BibitemOpen
  \bibfield  {author} {\bibinfo {author} {\bibfnamefont {G.}~\bibnamefont
  {Li}}, \bibinfo {author} {\bibfnamefont {Z.}~\bibnamefont {Xiang}}, \bibinfo
  {author} {\bibfnamefont {F.}~\bibnamefont {Yu}}, \bibinfo {author}
  {\bibfnamefont {T.}~\bibnamefont {Asaba}}, \bibinfo {author} {\bibfnamefont
  {B.}~\bibnamefont {Lawson}}, \bibinfo {author} {\bibfnamefont
  {P.}~\bibnamefont {Cai}}, \bibinfo {author} {\bibfnamefont {C.}~\bibnamefont
  {Tinsman}}, \bibinfo {author} {\bibfnamefont {A.}~\bibnamefont {Berkley}},
  \bibinfo {author} {\bibfnamefont {S.}~\bibnamefont {Wolgast}}, \bibinfo
  {author} {\bibfnamefont {Y.~S.}\ \bibnamefont {Eo}}, \bibinfo {author}
  {\bibfnamefont {D.-J.}\ \bibnamefont {Kim}}, \bibinfo {author} {\bibfnamefont
  {C.}~\bibnamefont {Kurdak}}, \bibinfo {author} {\bibfnamefont {J.~W.}\
  \bibnamefont {Allen}}, \bibinfo {author} {\bibfnamefont {K.}~\bibnamefont
  {Sun}}, \bibinfo {author} {\bibfnamefont {X.~H.}\ \bibnamefont {Chen}},
  \bibinfo {author} {\bibfnamefont {Y.~Y.}\ \bibnamefont {Wang}}, \bibinfo
  {author} {\bibfnamefont {Z.}~\bibnamefont {Fisk}}, \ and\ \bibinfo {author}
  {\bibfnamefont {L.}~\bibnamefont {Li}},\ }\bibfield  {title} {\enquote
  {\bibinfo {title} {Two-dimensional fermi surfaces in kondo insulator smb6},}\
  }\href {\doibase 10.1126/science.1250366} {\bibfield  {journal} {\bibinfo
  {journal} {Science}\ }\textbf {\bibinfo {volume} {346}},\ \bibinfo {pages}
  {1208} (\bibinfo {year} {2014})},\ \Eprint
  {http://arxiv.org/abs/http://science.sciencemag.org/content/346/6214/1208.full.pdf}
  {http://science.sciencemag.org/content/346/6214/1208.full.pdf} \BibitemShut
  {NoStop}%
\bibitem [{\citenamefont {{Denlinger}}\ \emph {et~al.}(2016)\citenamefont
  {{Denlinger}}, \citenamefont {{Jang}}, \citenamefont {{Li}}, \citenamefont
  {{Chen}}, \citenamefont {{Lawson}}, \citenamefont {{Asaba}}, \citenamefont
  {{Tinsman}}, \citenamefont {{Yu}}, \citenamefont {{Sun}}, \citenamefont
  {{Allen}}, \citenamefont {{Kurdak}}, \citenamefont {{Kim}}, \citenamefont
  {{Fisk}},\ and\ \citenamefont {{Li}}}]{Li16}%
  \BibitemOpen
  \bibfield  {author} {\bibinfo {author} {\bibfnamefont {J.~D.}\ \bibnamefont
  {{Denlinger}}}, \bibinfo {author} {\bibfnamefont {S.}~\bibnamefont {{Jang}}},
  \bibinfo {author} {\bibfnamefont {G.}~\bibnamefont {{Li}}}, \bibinfo {author}
  {\bibfnamefont {L.}~\bibnamefont {{Chen}}}, \bibinfo {author} {\bibfnamefont
  {B.~J.}\ \bibnamefont {{Lawson}}}, \bibinfo {author} {\bibfnamefont
  {T.}~\bibnamefont {{Asaba}}}, \bibinfo {author} {\bibfnamefont
  {C.}~\bibnamefont {{Tinsman}}}, \bibinfo {author} {\bibfnamefont
  {F.}~\bibnamefont {{Yu}}}, \bibinfo {author} {\bibfnamefont {K.}~\bibnamefont
  {{Sun}}}, \bibinfo {author} {\bibfnamefont {J.~W.}\ \bibnamefont {{Allen}}},
  \bibinfo {author} {\bibfnamefont {C.}~\bibnamefont {{Kurdak}}}, \bibinfo
  {author} {\bibfnamefont {D.-J.}\ \bibnamefont {{Kim}}}, \bibinfo {author}
  {\bibfnamefont {Z.}~\bibnamefont {{Fisk}}}, \ and\ \bibinfo {author}
  {\bibfnamefont {L.}~\bibnamefont {{Li}}},\ }\bibfield  {title} {\enquote
  {\bibinfo {title} {{Consistency of Photoemission and Quantum Oscillations for
  Surface States of SmB6}},}\ }\href@noop {} {\bibfield  {journal} {\bibinfo
  {journal} {ArXiv e-prints}\ } (\bibinfo {year} {2016})},\ \Eprint
  {http://arxiv.org/abs/1601.07408} {arXiv:1601.07408 [cond-mat.str-el]}
  \BibitemShut {NoStop}%
\bibitem [{\citenamefont {Tan}\ \emph {et~al.}(2015)\citenamefont {Tan},
  \citenamefont {Hsu}, \citenamefont {Zeng}, \citenamefont {Hatnean},
  \citenamefont {Harrison}, \citenamefont {Zhu}, \citenamefont {Hartstein},
  \citenamefont {Kiourlappou}, \citenamefont {Srivastava}, \citenamefont
  {Johannes}, \citenamefont {Murphy}, \citenamefont {Park}, \citenamefont
  {Balicas}, \citenamefont {Lonzarich}, \citenamefont {Balakrishnan},\ and\
  \citenamefont {Sebastian}}]{SS15}%
  \BibitemOpen
  \bibfield  {author} {\bibinfo {author} {\bibfnamefont {B.~S.}\ \bibnamefont
  {Tan}}, \bibinfo {author} {\bibfnamefont {Y.-T.}\ \bibnamefont {Hsu}},
  \bibinfo {author} {\bibfnamefont {B.}~\bibnamefont {Zeng}}, \bibinfo {author}
  {\bibfnamefont {M.~C.}\ \bibnamefont {Hatnean}}, \bibinfo {author}
  {\bibfnamefont {N.}~\bibnamefont {Harrison}}, \bibinfo {author}
  {\bibfnamefont {Z.}~\bibnamefont {Zhu}}, \bibinfo {author} {\bibfnamefont
  {M.}~\bibnamefont {Hartstein}}, \bibinfo {author} {\bibfnamefont
  {M.}~\bibnamefont {Kiourlappou}}, \bibinfo {author} {\bibfnamefont
  {A.}~\bibnamefont {Srivastava}}, \bibinfo {author} {\bibfnamefont {M.~D.}\
  \bibnamefont {Johannes}}, \bibinfo {author} {\bibfnamefont {T.~P.}\
  \bibnamefont {Murphy}}, \bibinfo {author} {\bibfnamefont {J.-H.}\
  \bibnamefont {Park}}, \bibinfo {author} {\bibfnamefont {L.}~\bibnamefont
  {Balicas}}, \bibinfo {author} {\bibfnamefont {G.~G.}\ \bibnamefont
  {Lonzarich}}, \bibinfo {author} {\bibfnamefont {G.}~\bibnamefont
  {Balakrishnan}}, \ and\ \bibinfo {author} {\bibfnamefont {S.~E.}\
  \bibnamefont {Sebastian}},\ }\bibfield  {title} {\enquote {\bibinfo {title}
  {Unconventional fermi surface in an insulating state},}\ }\href {\doibase
  10.1126/science.aaa7974} {\bibfield  {journal} {\bibinfo  {journal}
  {Science}\ }\textbf {\bibinfo {volume} {349}},\ \bibinfo {pages} {287}
  (\bibinfo {year} {2015})},\ \Eprint
  {http://arxiv.org/abs/http://science.sciencemag.org/content/349/6245/287.full.pdf}
  {http://science.sciencemag.org/content/349/6245/287.full.pdf} \BibitemShut
  {NoStop}%
\bibitem [{\citenamefont {Hartstein}\ \emph {et~al.}(2017)\citenamefont
  {Hartstein}, \citenamefont {Toews}, \citenamefont {Hsu},\ and\ \citenamefont
  {{\it et al.}}}]{SSnew}%
  \BibitemOpen
  \bibfield  {author} {\bibinfo {author} {\bibfnamefont {M.}~\bibnamefont
  {Hartstein}}, \bibinfo {author} {\bibfnamefont {W.}~\bibnamefont {Toews}},
  \bibinfo {author} {\bibfnamefont {Y.-T.}\ \bibnamefont {Hsu}}, \ and\
  \bibinfo {author} {\bibnamefont {{\it et al.}}},\ }\href@noop {} {\bibfield
  {journal} {\bibinfo  {journal} {{\it Unpublished}; S. Sebastian, private
  communication}\ } (\bibinfo {year} {2017})}\BibitemShut {NoStop}%
\bibitem [{\citenamefont {{Flachbart}}\ \emph {et~al.}(2006)\citenamefont
  {{Flachbart}}, \citenamefont {{Gab{\'a}ni}}, \citenamefont {{Neumaier}},
  \citenamefont {{Paderno}}, \citenamefont {{Pavl{\'{\i}}k}}, \citenamefont
  {{Schuberth}},\ and\ \citenamefont {{Shitsevalova}}}]{flachbart}%
  \BibitemOpen
  \bibfield  {author} {\bibinfo {author} {\bibfnamefont {K.}~\bibnamefont
  {{Flachbart}}}, \bibinfo {author} {\bibfnamefont {S.}~\bibnamefont
  {{Gab{\'a}ni}}}, \bibinfo {author} {\bibfnamefont {K.}~\bibnamefont
  {{Neumaier}}}, \bibinfo {author} {\bibfnamefont {Y.}~\bibnamefont
  {{Paderno}}}, \bibinfo {author} {\bibfnamefont {V.}~\bibnamefont
  {{Pavl{\'{\i}}k}}}, \bibinfo {author} {\bibfnamefont {E.}~\bibnamefont
  {{Schuberth}}}, \ and\ \bibinfo {author} {\bibfnamefont {N.}~\bibnamefont
  {{Shitsevalova}}},\ }\bibfield  {title} {\enquote {\bibinfo {title}
  {{Specific heat of SmB $_{6}$ at very low temperatures}},}\ }\href {\doibase
  10.1016/j.physb.2006.01.344} {\bibfield  {journal} {\bibinfo  {journal}
  {Physica B Condensed Matter}\ }\textbf {\bibinfo {volume} {378}},\ \bibinfo
  {pages} {610} (\bibinfo {year} {2006})}\BibitemShut {NoStop}%
\bibitem [{\citenamefont {Wakeham}\ \emph {et~al.}(2016)\citenamefont
  {Wakeham}, \citenamefont {Rosa}, \citenamefont {Wang}, \citenamefont {Kang},
  \citenamefont {Fisk}, \citenamefont {Ronning},\ and\ \citenamefont
  {Thompson}}]{thompson}%
  \BibitemOpen
  \bibfield  {author} {\bibinfo {author} {\bibfnamefont {N.}~\bibnamefont
  {Wakeham}}, \bibinfo {author} {\bibfnamefont {P.~F.~S.}\ \bibnamefont
  {Rosa}}, \bibinfo {author} {\bibfnamefont {Y.~Q.}\ \bibnamefont {Wang}},
  \bibinfo {author} {\bibfnamefont {M.}~\bibnamefont {Kang}}, \bibinfo {author}
  {\bibfnamefont {Z.}~\bibnamefont {Fisk}}, \bibinfo {author} {\bibfnamefont
  {F.}~\bibnamefont {Ronning}}, \ and\ \bibinfo {author} {\bibfnamefont
  {J.~D.}\ \bibnamefont {Thompson}},\ }\bibfield  {title} {\enquote {\bibinfo
  {title} {Low-temperature conducting state in two candidate topological kondo
  insulators: ${\mathrm{smb}}_{6}$ and
  ${\mathrm{ce}}_{3}{\mathrm{bi}}_{4}{\mathrm{pt}}_{3}$},}\ }\href {\doibase
  10.1103/PhysRevB.94.035127} {\bibfield  {journal} {\bibinfo  {journal} {Phys.
  Rev. B}\ }\textbf {\bibinfo {volume} {94}},\ \bibinfo {pages} {035127}
  (\bibinfo {year} {2016})}\BibitemShut {NoStop}%
\bibitem [{\citenamefont {Laurita}\ \emph {et~al.}(2016)\citenamefont
  {Laurita}, \citenamefont {Morris}, \citenamefont {Koohpayeh}, \citenamefont
  {Rosa}, \citenamefont {Phelan}, \citenamefont {Fisk}, \citenamefont
  {McQueen},\ and\ \citenamefont {Armitage}}]{Armitage16}%
  \BibitemOpen
  \bibfield  {author} {\bibinfo {author} {\bibfnamefont {N.~J.}\ \bibnamefont
  {Laurita}}, \bibinfo {author} {\bibfnamefont {C.~M.}\ \bibnamefont {Morris}},
  \bibinfo {author} {\bibfnamefont {S.~M.}\ \bibnamefont {Koohpayeh}}, \bibinfo
  {author} {\bibfnamefont {P.~F.~S.}\ \bibnamefont {Rosa}}, \bibinfo {author}
  {\bibfnamefont {W.~A.}\ \bibnamefont {Phelan}}, \bibinfo {author}
  {\bibfnamefont {Z.}~\bibnamefont {Fisk}}, \bibinfo {author} {\bibfnamefont
  {T.~M.}\ \bibnamefont {McQueen}}, \ and\ \bibinfo {author} {\bibfnamefont
  {N.~P.}\ \bibnamefont {Armitage}},\ }\bibfield  {title} {\enquote {\bibinfo
  {title} {Anomalous three-dimensional bulk ac conduction within the kondo gap
  of ${\mathrm{smb}}_{6}$ single crystals},}\ }\href {\doibase
  10.1103/PhysRevB.94.165154} {\bibfield  {journal} {\bibinfo  {journal} {Phys.
  Rev. B}\ }\textbf {\bibinfo {volume} {94}},\ \bibinfo {pages} {165154}
  (\bibinfo {year} {2016})}\BibitemShut {NoStop}%
\bibitem [{\citenamefont {{Chowdhury}}\ \emph {et~al.}(2017)\citenamefont
  {{Chowdhury}}, \citenamefont {{Sodemann}},\ and\ \citenamefont
  {{Senthil}}}]{CSS}%
  \BibitemOpen
  \bibfield  {author} {\bibinfo {author} {\bibfnamefont {D.}~\bibnamefont
  {{Chowdhury}}}, \bibinfo {author} {\bibfnamefont {I.}~\bibnamefont
  {{Sodemann}}}, \ and\ \bibinfo {author} {\bibfnamefont {T.}~\bibnamefont
  {{Senthil}}},\ }\bibfield  {title} {\enquote {\bibinfo {title}
  {{Mixed-valence insulators with neutral Fermi-surfaces}},}\ }\href@noop {}
  {\bibfield  {journal} {\bibinfo  {journal} {ArXiv e-prints}\ } (\bibinfo
  {year} {2017})},\ \Eprint {http://arxiv.org/abs/1706.00418} {arXiv:1706.00418
  [cond-mat.str-el]} \BibitemShut {NoStop}%
\bibitem [{\citenamefont {{Baskaran}}(2015)}]{Baskaran}%
  \BibitemOpen
  \bibfield  {author} {\bibinfo {author} {\bibfnamefont {G.}~\bibnamefont
  {{Baskaran}}},\ }\bibfield  {title} {\enquote {\bibinfo {title} {{Majorana
  Fermi Sea in Insulating SmB6: A proposal and a Theory of Quantum Oscillations
  in Kondo Insulators}},}\ }\href@noop {} {\bibfield  {journal} {\bibinfo
  {journal} {ArXiv e-prints}\ } (\bibinfo {year} {2015})},\ \Eprint
  {http://arxiv.org/abs/1507.03477} {arXiv:1507.03477 [cond-mat.str-el]}
  \BibitemShut {NoStop}%
\bibitem [{\citenamefont {{Erten}}\ \emph {et~al.}(2017)\citenamefont
  {{Erten}}, \citenamefont {{Chang}}, \citenamefont {{Coleman}},\ and\
  \citenamefont {{Tsvelik}}}]{ColemanSC}%
  \BibitemOpen
  \bibfield  {author} {\bibinfo {author} {\bibfnamefont {O.}~\bibnamefont
  {{Erten}}}, \bibinfo {author} {\bibfnamefont {P.-Y.}\ \bibnamefont
  {{Chang}}}, \bibinfo {author} {\bibfnamefont {P.}~\bibnamefont {{Coleman}}},
  \ and\ \bibinfo {author} {\bibfnamefont {A.~M.}\ \bibnamefont {{Tsvelik}}},\
  }\bibfield  {title} {\enquote {\bibinfo {title} {{Skyrme insulators:
  insulators at the brink of superconductivity}},}\ }\href@noop {} {\bibfield
  {journal} {\bibinfo  {journal} {ArXiv e-prints}\ } (\bibinfo {year}
  {2017})},\ \Eprint {http://arxiv.org/abs/1701.06582} {arXiv:1701.06582
  [cond-mat.str-el]} \BibitemShut {NoStop}%
\bibitem [{\citenamefont {Knolle}\ and\ \citenamefont
  {Cooper}(2017)}]{CooperExc}%
  \BibitemOpen
  \bibfield  {author} {\bibinfo {author} {\bibfnamefont {J.}~\bibnamefont
  {Knolle}}\ and\ \bibinfo {author} {\bibfnamefont {N.~R.}\ \bibnamefont
  {Cooper}},\ }\bibfield  {title} {\enquote {\bibinfo {title} {Excitons in
  topological kondo insulators: Theory of thermodynamic and transport anomalies
  in ${\mathrm{smb}}_{6}$},}\ }\href {\doibase 10.1103/PhysRevLett.118.096604}
  {\bibfield  {journal} {\bibinfo  {journal} {Phys. Rev. Lett.}\ }\textbf
  {\bibinfo {volume} {118}},\ \bibinfo {pages} {096604} (\bibinfo {year}
  {2017})}\BibitemShut {NoStop}%
\bibitem [{\citenamefont {Knolle}\ and\ \citenamefont
  {Cooper}(2015)}]{CooperMB}%
  \BibitemOpen
  \bibfield  {author} {\bibinfo {author} {\bibfnamefont {J.}~\bibnamefont
  {Knolle}}\ and\ \bibinfo {author} {\bibfnamefont {N.~R.}\ \bibnamefont
  {Cooper}},\ }\bibfield  {title} {\enquote {\bibinfo {title} {Quantum
  oscillations without a fermi surface and the anomalous de haas\char21{}van
  alphen effect},}\ }\href {\doibase 10.1103/PhysRevLett.115.146401} {\bibfield
   {journal} {\bibinfo  {journal} {Phys. Rev. Lett.}\ }\textbf {\bibinfo
  {volume} {115}},\ \bibinfo {pages} {146401} (\bibinfo {year}
  {2015})}\BibitemShut {NoStop}%
\bibitem [{\citenamefont {Zhang}\ \emph {et~al.}(2016)\citenamefont {Zhang},
  \citenamefont {Song},\ and\ \citenamefont {Wang}}]{FWMB}%
  \BibitemOpen
  \bibfield  {author} {\bibinfo {author} {\bibfnamefont {L.}~\bibnamefont
  {Zhang}}, \bibinfo {author} {\bibfnamefont {X.-Y.}\ \bibnamefont {Song}}, \
  and\ \bibinfo {author} {\bibfnamefont {F.}~\bibnamefont {Wang}},\ }\bibfield
  {title} {\enquote {\bibinfo {title} {Quantum oscillation in narrow-gap
  topological insulators},}\ }\href {\doibase 10.1103/PhysRevLett.116.046404}
  {\bibfield  {journal} {\bibinfo  {journal} {Phys. Rev. Lett.}\ }\textbf
  {\bibinfo {volume} {116}},\ \bibinfo {pages} {046404} (\bibinfo {year}
  {2016})}\BibitemShut {NoStop}%
\bibitem [{\citenamefont {Ram}\ and\ \citenamefont {Kumar}(2017)}]{Kumar2017}%
  \BibitemOpen
  \bibfield  {author} {\bibinfo {author} {\bibfnamefont {P.}~\bibnamefont
  {Ram}}\ and\ \bibinfo {author} {\bibfnamefont {B.}~\bibnamefont {Kumar}},\
  }\bibfield  {title} {\enquote {\bibinfo {title} {Theory of quantum
  oscillations of magnetization in kondo insulators},}\ }\href {\doibase
  10.1103/PhysRevB.96.075115} {\bibfield  {journal} {\bibinfo  {journal} {Phys.
  Rev. B}\ }\textbf {\bibinfo {volume} {96}},\ \bibinfo {pages} {075115}
  (\bibinfo {year} {2017})}\BibitemShut {NoStop}%
\bibitem [{\citenamefont {Read}(1994)}]{NR94}%
  \BibitemOpen
  \bibfield  {author} {\bibinfo {author} {\bibfnamefont {N.}~\bibnamefont
  {Read}},\ }\bibfield  {title} {\enquote {\bibinfo {title} {Theory of the
  half-filled landau level},}\ }\href
  {http://stacks.iop.org/0268-1242/9/i=11S/a=002} {\bibfield  {journal}
  {\bibinfo  {journal} {Semiconductor Science and Technology}\ }\textbf
  {\bibinfo {volume} {9}},\ \bibinfo {pages} {1859} (\bibinfo {year}
  {1994})}\BibitemShut {NoStop}%
\bibitem [{\citenamefont {Read}(1998)}]{NR98}%
  \BibitemOpen
  \bibfield  {author} {\bibinfo {author} {\bibfnamefont {N.}~\bibnamefont
  {Read}},\ }\bibfield  {title} {\enquote {\bibinfo {title}
  {Lowest-landau-level theory of the quantum hall effect: The fermi-liquid-like
  state of bosons at filling factor one},}\ }\href {\doibase
  10.1103/PhysRevB.58.16262} {\bibfield  {journal} {\bibinfo  {journal} {Phys.
  Rev. B}\ }\textbf {\bibinfo {volume} {58}},\ \bibinfo {pages} {16262}
  (\bibinfo {year} {1998})}\BibitemShut {NoStop}%
\bibitem [{\citenamefont {Halperin}\ \emph {et~al.}(1993)\citenamefont
  {Halperin}, \citenamefont {Lee},\ and\ \citenamefont {Read}}]{HLR}%
  \BibitemOpen
  \bibfield  {author} {\bibinfo {author} {\bibfnamefont {B.~I.}\ \bibnamefont
  {Halperin}}, \bibinfo {author} {\bibfnamefont {P.~A.}\ \bibnamefont {Lee}}, \
  and\ \bibinfo {author} {\bibfnamefont {N.}~\bibnamefont {Read}},\ }\bibfield
  {title} {\enquote {\bibinfo {title} {Theory of the half-filled landau
  level},}\ }\href {\doibase 10.1103/PhysRevB.47.7312} {\bibfield  {journal}
  {\bibinfo  {journal} {Phys. Rev. B}\ }\textbf {\bibinfo {volume} {47}},\
  \bibinfo {pages} {7312} (\bibinfo {year} {1993})}\BibitemShut {NoStop}%
\bibitem [{\citenamefont {Abrikosov}\ and\ \citenamefont
  {Beknazarov}(1988)}]{abrikosov}%
  \BibitemOpen
  \bibfield  {author} {\bibinfo {author} {\bibfnamefont {A.~A.}\ \bibnamefont
  {Abrikosov}}\ and\ \bibinfo {author} {\bibfnamefont {A.}~\bibnamefont
  {Beknazarov}},\ }\href@noop {} {\emph {\bibinfo {title} {Fundamentals of the
  Theory of Metals}}},\ Vol.~\bibinfo {volume} {1}\ (\bibinfo  {publisher}
  {North-Holland Amsterdam},\ \bibinfo {year} {1988})\BibitemShut {NoStop}%
\bibitem [{\citenamefont {Shoenberg}(2009)}]{DS}%
  \BibitemOpen
  \bibfield  {author} {\bibinfo {author} {\bibfnamefont {D.}~\bibnamefont
  {Shoenberg}},\ }\href@noop {} {\emph {\bibinfo {title} {Magnetic oscillations
  in metals}}}\ (\bibinfo  {publisher} {Cambridge University Press},\ \bibinfo
  {year} {2009})\BibitemShut {NoStop}%
\bibitem [{\citenamefont {Solt}\ \emph {et~al.}(1999)\citenamefont {Solt},
  \citenamefont {Baines}, \citenamefont {Egorov}, \citenamefont {Herlach},\
  and\ \citenamefont {Zimmermann}}]{Domains99}%
  \BibitemOpen
  \bibfield  {author} {\bibinfo {author} {\bibfnamefont {G.}~\bibnamefont
  {Solt}}, \bibinfo {author} {\bibfnamefont {C.}~\bibnamefont {Baines}},
  \bibinfo {author} {\bibfnamefont {V.~S.}\ \bibnamefont {Egorov}}, \bibinfo
  {author} {\bibfnamefont {D.}~\bibnamefont {Herlach}}, \ and\ \bibinfo
  {author} {\bibfnamefont {U.}~\bibnamefont {Zimmermann}},\ }\bibfield  {title}
  {\enquote {\bibinfo {title} {Diamagnetic domains in beryllium observed by
  muon-spin-rotation spectroscopy},}\ }\href {\doibase
  10.1103/PhysRevB.59.6834} {\bibfield  {journal} {\bibinfo  {journal} {Phys.
  Rev. B}\ }\textbf {\bibinfo {volume} {59}},\ \bibinfo {pages} {6834}
  (\bibinfo {year} {1999})}\BibitemShut {NoStop}%
\bibitem [{\citenamefont {Gordon}\ \emph {et~al.}(1999)\citenamefont {Gordon},
  \citenamefont {Itskovsky},\ and\ \citenamefont {Wyder}}]{Domains99b}%
  \BibitemOpen
  \bibfield  {author} {\bibinfo {author} {\bibfnamefont {A.}~\bibnamefont
  {Gordon}}, \bibinfo {author} {\bibfnamefont {M.~A.}\ \bibnamefont
  {Itskovsky}}, \ and\ \bibinfo {author} {\bibfnamefont {P.}~\bibnamefont
  {Wyder}},\ }\bibfield  {title} {\enquote {\bibinfo {title} {Quantizing
  field-induced magnetic phase in a three-dimensional electron gas},}\ }\href
  {\doibase 10.1103/PhysRevB.59.10864} {\bibfield  {journal} {\bibinfo
  {journal} {Phys. Rev. B}\ }\textbf {\bibinfo {volume} {59}},\ \bibinfo
  {pages} {10864} (\bibinfo {year} {1999})}\BibitemShut {NoStop}%
\bibitem [{\citenamefont {Gordon}\ \emph {et~al.}(2003)\citenamefont {Gordon},
  \citenamefont {Vagner},\ and\ \citenamefont {Wyder}}]{gordon2003magnetic}%
  \BibitemOpen
  \bibfield  {author} {\bibinfo {author} {\bibfnamefont {A.}~\bibnamefont
  {Gordon}}, \bibinfo {author} {\bibfnamefont {I.}~\bibnamefont {Vagner}}, \
  and\ \bibinfo {author} {\bibfnamefont {P.}~\bibnamefont {Wyder}},\ }\bibfield
   {title} {\enquote {\bibinfo {title} {Magnetic domains in non-ferromagnetic
  metals: the non-linear de haas-van alphen effect},}\ }\href@noop {}
  {\bibfield  {journal} {\bibinfo  {journal} {Advances in Physics}\ }\textbf
  {\bibinfo {volume} {52}},\ \bibinfo {pages} {385} (\bibinfo {year}
  {2003})}\BibitemShut {NoStop}%
\bibitem [{\citenamefont {Kramer}\ \emph {et~al.}(2005)\citenamefont {Kramer},
  \citenamefont {Egorov}, \citenamefont {Gasparov}, \citenamefont {Jansen},\
  and\ \citenamefont {Joss}}]{Domains05}%
  \BibitemOpen
  \bibfield  {author} {\bibinfo {author} {\bibfnamefont {R.~B.~G.}\
  \bibnamefont {Kramer}}, \bibinfo {author} {\bibfnamefont {V.~S.}\
  \bibnamefont {Egorov}}, \bibinfo {author} {\bibfnamefont {V.~A.}\
  \bibnamefont {Gasparov}}, \bibinfo {author} {\bibfnamefont {A.~G.~M.}\
  \bibnamefont {Jansen}}, \ and\ \bibinfo {author} {\bibfnamefont
  {W.}~\bibnamefont {Joss}},\ }\bibfield  {title} {\enquote {\bibinfo {title}
  {Direct observation of condon domains in silver by hall probes},}\ }\href
  {\doibase 10.1103/PhysRevLett.95.267209} {\bibfield  {journal} {\bibinfo
  {journal} {Phys. Rev. Lett.}\ }\textbf {\bibinfo {volume} {95}},\ \bibinfo
  {pages} {267209} (\bibinfo {year} {2005})}\BibitemShut {NoStop}%
\bibitem [{\citenamefont {Egorov}(2005)}]{egorov2005condon}%
  \BibitemOpen
  \bibfield  {author} {\bibinfo {author} {\bibfnamefont {V.~S.}\ \bibnamefont
  {Egorov}},\ }\bibfield  {title} {\enquote {\bibinfo {title} {Condon
  domains-these non-magnetic diamagnetic domains},}\ }\href@noop {} {\bibfield
  {journal} {\bibinfo  {journal} {arXiv preprint cond-mat/0505415}\ } (\bibinfo
  {year} {2005})}\BibitemShut {NoStop}%
\bibitem [{\citenamefont {Hermele}\ \emph {et~al.}(2004)\citenamefont
  {Hermele}, \citenamefont {Senthil}, \citenamefont {Fisher}, \citenamefont
  {Lee}, \citenamefont {Nagaosa},\ and\ \citenamefont {Wen}}]{monopoles1}%
  \BibitemOpen
  \bibfield  {author} {\bibinfo {author} {\bibfnamefont {M.}~\bibnamefont
  {Hermele}}, \bibinfo {author} {\bibfnamefont {T.}~\bibnamefont {Senthil}},
  \bibinfo {author} {\bibfnamefont {M.~P.~A.}\ \bibnamefont {Fisher}}, \bibinfo
  {author} {\bibfnamefont {P.~A.}\ \bibnamefont {Lee}}, \bibinfo {author}
  {\bibfnamefont {N.}~\bibnamefont {Nagaosa}}, \ and\ \bibinfo {author}
  {\bibfnamefont {X.-G.}\ \bibnamefont {Wen}},\ }\bibfield  {title} {\enquote
  {\bibinfo {title} {Stability of $u(1)$ spin liquids in two dimensions},}\
  }\href {\doibase 10.1103/PhysRevB.70.214437} {\bibfield  {journal} {\bibinfo
  {journal} {Phys. Rev. B}\ }\textbf {\bibinfo {volume} {70}},\ \bibinfo
  {pages} {214437} (\bibinfo {year} {2004})}\BibitemShut {NoStop}%
\bibitem [{\citenamefont {Lee}(2008)}]{monopoles2}%
  \BibitemOpen
  \bibfield  {author} {\bibinfo {author} {\bibfnamefont {S.-S.}\ \bibnamefont
  {Lee}},\ }\bibfield  {title} {\enquote {\bibinfo {title} {Stability of the
  u(1) spin liquid with a spinon fermi surface in $2+1$ dimensions},}\ }\href
  {\doibase 10.1103/PhysRevB.78.085129} {\bibfield  {journal} {\bibinfo
  {journal} {Phys. Rev. B}\ }\textbf {\bibinfo {volume} {78}},\ \bibinfo
  {pages} {085129} (\bibinfo {year} {2008})}\BibitemShut {NoStop}%
\bibitem [{\citenamefont {Florens}\ and\ \citenamefont
  {Georges}(2004)}]{FG2004}%
  \BibitemOpen
  \bibfield  {author} {\bibinfo {author} {\bibfnamefont {S.}~\bibnamefont
  {Florens}}\ and\ \bibinfo {author} {\bibfnamefont {A.}~\bibnamefont
  {Georges}},\ }\bibfield  {title} {\enquote {\bibinfo {title} {Slave-rotor
  mean-field theories of strongly correlated systems and the mott transition in
  finite dimensions},}\ }\href {\doibase 10.1103/PhysRevB.70.035114} {\bibfield
   {journal} {\bibinfo  {journal} {Phys. Rev. B}\ }\textbf {\bibinfo {volume}
  {70}},\ \bibinfo {pages} {035114} (\bibinfo {year} {2004})}\BibitemShut
  {NoStop}%
\bibitem [{\citenamefont {Senthil}(2008)}]{TS2008}%
  \BibitemOpen
  \bibfield  {author} {\bibinfo {author} {\bibfnamefont {T.}~\bibnamefont
  {Senthil}},\ }\bibfield  {title} {\enquote {\bibinfo {title} {Theory of a
  continuous mott transition in two dimensions},}\ }\href {\doibase
  10.1103/PhysRevB.78.045109} {\bibfield  {journal} {\bibinfo  {journal} {Phys.
  Rev. B}\ }\textbf {\bibinfo {volume} {78}},\ \bibinfo {pages} {045109}
  (\bibinfo {year} {2008})}\BibitemShut {NoStop}%
\bibitem [{\citenamefont {Ioffe}\ and\ \citenamefont {Larkin}(1989)}]{IL}%
  \BibitemOpen
  \bibfield  {author} {\bibinfo {author} {\bibfnamefont {L.~B.}\ \bibnamefont
  {Ioffe}}\ and\ \bibinfo {author} {\bibfnamefont {A.~I.}\ \bibnamefont
  {Larkin}},\ }\bibfield  {title} {\enquote {\bibinfo {title} {Gapless fermions
  and gauge fields in dielectrics},}\ }\href {\doibase
  10.1103/PhysRevB.39.8988} {\bibfield  {journal} {\bibinfo  {journal} {Phys.
  Rev. B}\ }\textbf {\bibinfo {volume} {39}},\ \bibinfo {pages} {8988}
  (\bibinfo {year} {1989})}\BibitemShut {NoStop}%
\bibitem [{\citenamefont {Imry}\ and\ \citenamefont {Ma}(1975)}]{ImryMa}%
  \BibitemOpen
  \bibfield  {author} {\bibinfo {author} {\bibfnamefont {Y.}~\bibnamefont
  {Imry}}\ and\ \bibinfo {author} {\bibfnamefont {S.-k.}\ \bibnamefont {Ma}},\
  }\bibfield  {title} {\enquote {\bibinfo {title} {Random-field instability of
  the ordered state of continuous symmetry},}\ }\href {\doibase
  10.1103/PhysRevLett.35.1399} {\bibfield  {journal} {\bibinfo  {journal}
  {Phys. Rev. Lett.}\ }\textbf {\bibinfo {volume} {35}},\ \bibinfo {pages}
  {1399} (\bibinfo {year} {1975})}\BibitemShut {NoStop}%
\bibitem [{\citenamefont {Imry}\ and\ \citenamefont
  {Wortis}(1979)}]{ImryWortis}%
  \BibitemOpen
  \bibfield  {author} {\bibinfo {author} {\bibfnamefont {Y.}~\bibnamefont
  {Imry}}\ and\ \bibinfo {author} {\bibfnamefont {M.}~\bibnamefont {Wortis}},\
  }\bibfield  {title} {\enquote {\bibinfo {title} {Influence of quenched
  impurities on first-order phase transitions},}\ }\href {\doibase
  10.1103/PhysRevB.19.3580} {\bibfield  {journal} {\bibinfo  {journal} {Phys.
  Rev. B}\ }\textbf {\bibinfo {volume} {19}},\ \bibinfo {pages} {3580}
  (\bibinfo {year} {1979})}\BibitemShut {NoStop}%
\bibitem [{\citenamefont {Aizenman}\ and\ \citenamefont
  {Wehr}(1989)}]{AizenmanWehr}%
  \BibitemOpen
  \bibfield  {author} {\bibinfo {author} {\bibfnamefont {M.}~\bibnamefont
  {Aizenman}}\ and\ \bibinfo {author} {\bibfnamefont {J.}~\bibnamefont
  {Wehr}},\ }\bibfield  {title} {\enquote {\bibinfo {title} {Rounding of
  first-order phase transitions in systems with quenched disorder},}\ }\href
  {\doibase 10.1103/PhysRevLett.62.2503} {\bibfield  {journal} {\bibinfo
  {journal} {Phys. Rev. Lett.}\ }\textbf {\bibinfo {volume} {62}},\ \bibinfo
  {pages} {2503} (\bibinfo {year} {1989})}\BibitemShut {NoStop}%
\bibitem [{\citenamefont {Dzero}\ \emph {et~al.}(2010)\citenamefont {Dzero},
  \citenamefont {Sun}, \citenamefont {Galitski},\ and\ \citenamefont
  {Coleman}}]{Coleman1}%
  \BibitemOpen
  \bibfield  {author} {\bibinfo {author} {\bibfnamefont {M.}~\bibnamefont
  {Dzero}}, \bibinfo {author} {\bibfnamefont {K.}~\bibnamefont {Sun}}, \bibinfo
  {author} {\bibfnamefont {V.}~\bibnamefont {Galitski}}, \ and\ \bibinfo
  {author} {\bibfnamefont {P.}~\bibnamefont {Coleman}},\ }\bibfield  {title}
  {\enquote {\bibinfo {title} {Topological kondo insulators},}\ }\href
  {\doibase 10.1103/PhysRevLett.104.106408} {\bibfield  {journal} {\bibinfo
  {journal} {Phys. Rev. Lett.}\ }\textbf {\bibinfo {volume} {104}},\ \bibinfo
  {pages} {106408} (\bibinfo {year} {2010})}\BibitemShut {NoStop}%
\bibitem [{\citenamefont {Matsuda}\ \emph {et~al.}(2017)\citenamefont
  {Matsuda}, \citenamefont {Li},\ and\ \citenamefont {Shibauchi}}]{MatsudaLi}%
  \BibitemOpen
  \bibfield  {author} {\bibinfo {author} {\bibfnamefont {Y.}~\bibnamefont
  {Matsuda}}, \bibinfo {author} {\bibfnamefont {L.}~\bibnamefont {Li}}, \ and\
  \bibinfo {author} {\bibfnamefont {T.}~\bibnamefont {Shibauchi}},\ }\href@noop
  {} {\bibfield  {journal} {\bibinfo  {journal} {private communication}\ }
  (\bibinfo {year} {2017})}\BibitemShut {NoStop}%
\bibitem [{\citenamefont {Komatsu}\ \emph {et~al.}(1996)\citenamefont
  {Komatsu}, \citenamefont {Matsukawa}, \citenamefont {Inoue},\ and\
  \citenamefont {Saito}}]{pressure1996}%
  \BibitemOpen
  \bibfield  {author} {\bibinfo {author} {\bibfnamefont {T.}~\bibnamefont
  {Komatsu}}, \bibinfo {author} {\bibfnamefont {N.}~\bibnamefont {Matsukawa}},
  \bibinfo {author} {\bibfnamefont {T.}~\bibnamefont {Inoue}}, \ and\ \bibinfo
  {author} {\bibfnamefont {G.}~\bibnamefont {Saito}},\ }\bibfield  {title}
  {\enquote {\bibinfo {title} {Realization of superconductivity at ambient
  pressure by band-filling control in κ- (bedt-ttf)2cu2(cn)3},}\ }\href
  {\doibase 10.1143/JPSJ.65.1340} {\bibfield  {journal} {\bibinfo  {journal}
  {Journal of the Physical Society of Japan}\ }\textbf {\bibinfo {volume}
  {65}},\ \bibinfo {pages} {1340} (\bibinfo {year} {1996})},\ \Eprint
  {http://arxiv.org/abs/http://dx.doi.org/10.1143/JPSJ.65.1340}
  {http://dx.doi.org/10.1143/JPSJ.65.1340} \BibitemShut {NoStop}%
\bibitem [{\citenamefont {Kurosaki}\ \emph {et~al.}(2005)\citenamefont
  {Kurosaki}, \citenamefont {Shimizu}, \citenamefont {Miyagawa}, \citenamefont
  {Kanoda},\ and\ \citenamefont {Saito}}]{pressure2005}%
  \BibitemOpen
  \bibfield  {author} {\bibinfo {author} {\bibfnamefont {Y.}~\bibnamefont
  {Kurosaki}}, \bibinfo {author} {\bibfnamefont {Y.}~\bibnamefont {Shimizu}},
  \bibinfo {author} {\bibfnamefont {K.}~\bibnamefont {Miyagawa}}, \bibinfo
  {author} {\bibfnamefont {K.}~\bibnamefont {Kanoda}}, \ and\ \bibinfo {author}
  {\bibfnamefont {G.}~\bibnamefont {Saito}},\ }\bibfield  {title} {\enquote
  {\bibinfo {title} {Mott transition from a spin liquid to a fermi liquid in
  the spin-frustrated organic conductor
  $\ensuremath{\kappa}\mathrm{\text{\ensuremath{-}}}(\mathrm{ET}{)}_{2}{\mathrm{cu}}_{2}(\mathrm{CN}{)}_{3}$},}\
  }\href {\doibase 10.1103/PhysRevLett.95.177001} {\bibfield  {journal}
  {\bibinfo  {journal} {Phys. Rev. Lett.}\ }\textbf {\bibinfo {volume} {95}},\
  \bibinfo {pages} {177001} (\bibinfo {year} {2005})}\BibitemShut {NoStop}%
\bibitem [{\citenamefont {Furukawa}\ \emph {et~al.}(2015)\citenamefont
  {Furukawa}, \citenamefont {Miyagawa}, \citenamefont {Taniguchi},
  \citenamefont {Kato},\ and\ \citenamefont {Kanoda}}]{pressure2015}%
  \BibitemOpen
  \bibfield  {author} {\bibinfo {author} {\bibfnamefont {T.}~\bibnamefont
  {Furukawa}}, \bibinfo {author} {\bibfnamefont {K.}~\bibnamefont {Miyagawa}},
  \bibinfo {author} {\bibfnamefont {H.}~\bibnamefont {Taniguchi}}, \bibinfo
  {author} {\bibfnamefont {R.}~\bibnamefont {Kato}}, \ and\ \bibinfo {author}
  {\bibfnamefont {K.}~\bibnamefont {Kanoda}},\ }\bibfield  {title} {\enquote
  {\bibinfo {title} {Quantum criticality of mott transition in organic
  materials},}\ }\href@noop {} {\bibfield  {journal} {\bibinfo  {journal}
  {Nature Physics}\ }\textbf {\bibinfo {volume} {11}},\ \bibinfo {pages} {221}
  (\bibinfo {year} {2015})}\BibitemShut {NoStop}%
\bibitem [{\citenamefont {Furukawa}\ \emph {et~al.}(2017)\citenamefont
  {Furukawa}, \citenamefont {Kobashi}, \citenamefont {Kurosaki}, \citenamefont
  {Miyagawa},\ and\ \citenamefont {Kanoda}}]{Kanoda2017}%
  \BibitemOpen
  \bibfield  {author} {\bibinfo {author} {\bibfnamefont {T.}~\bibnamefont
  {Furukawa}}, \bibinfo {author} {\bibfnamefont {K.}~\bibnamefont {Kobashi}},
  \bibinfo {author} {\bibfnamefont {Y.}~\bibnamefont {Kurosaki}}, \bibinfo
  {author} {\bibfnamefont {K.}~\bibnamefont {Miyagawa}}, \ and\ \bibinfo
  {author} {\bibfnamefont {K.}~\bibnamefont {Kanoda}},\ }\bibfield  {title}
  {\enquote {\bibinfo {title} {Quasi-continuous transition from a fermi liquid
  to a spin liquid},}\ }\href@noop {} {\bibfield  {journal} {\bibinfo
  {journal} {arXiv preprint arXiv:1707.05586}\ } (\bibinfo {year}
  {2017})}\BibitemShut {NoStop}%
\bibitem [{\citenamefont {Watanabe}\ \emph {et~al.}(2012)\citenamefont
  {Watanabe}, \citenamefont {Yamashita}, \citenamefont {Tonegawa},
  \citenamefont {Oshima}, \citenamefont {Yamamoto}, \citenamefont {Kato},
  \citenamefont {Sheikin}, \citenamefont {Behnia}, \citenamefont {Terashima},
  \citenamefont {Uji} \emph {et~al.}}]{watanabe2012novel}%
  \BibitemOpen
  \bibfield  {author} {\bibinfo {author} {\bibfnamefont {D.}~\bibnamefont
  {Watanabe}}, \bibinfo {author} {\bibfnamefont {M.}~\bibnamefont {Yamashita}},
  \bibinfo {author} {\bibfnamefont {S.}~\bibnamefont {Tonegawa}}, \bibinfo
  {author} {\bibfnamefont {Y.}~\bibnamefont {Oshima}}, \bibinfo {author}
  {\bibfnamefont {H.}~\bibnamefont {Yamamoto}}, \bibinfo {author}
  {\bibfnamefont {R.}~\bibnamefont {Kato}}, \bibinfo {author} {\bibfnamefont
  {I.}~\bibnamefont {Sheikin}}, \bibinfo {author} {\bibfnamefont
  {K.}~\bibnamefont {Behnia}}, \bibinfo {author} {\bibfnamefont
  {T.}~\bibnamefont {Terashima}}, \bibinfo {author} {\bibfnamefont
  {S.}~\bibnamefont {Uji}},  \emph {et~al.},\ }\bibfield  {title} {\enquote
  {\bibinfo {title} {Novel pauli-paramagnetic quantum phase in a mott
  insulator},}\ }\href@noop {} {\bibfield  {journal} {\bibinfo  {journal}
  {Nature Communications}\ }\textbf {\bibinfo {volume} {3}},\ \bibinfo {pages}
  {1090} (\bibinfo {year} {2012})}\BibitemShut {NoStop}%
\end{thebibliography}

\end{document}